\documentclass[a4paper,11pt]{article}
\pdfoutput=1 

\usepackage{jheppub} 
\usepackage{xfrac}
\usepackage{caption}
\usepackage[dvipsnames]{xcolor}
\usepackage{relsize} 
\usepackage{pdfpages}
\usepackage{graphicx}
\usepackage{dcolumn}
\usepackage{bm}
\usepackage{epstopdf}
\usepackage{mathrsfs}
\usepackage{amsmath,amssymb,amsfonts,latexsym}
\usepackage{mathtools}
\usepackage{amsmath,bbold}
\usepackage{color}
\usepackage{slashed}
\usepackage{nicefrac}
\usepackage[colorlinks=true,linktocpage=true,linkcolor=blue,citecolor=blue]{hyperref}
\usepackage[utf8]{inputenc}
\usepackage[T1]{fontenc}
\usepackage{accents}
\usepackage[normalem]{ulem}

\long\def\comment#1{ }

\def\0{{\boldsymbol 0}}

\def\and{\qquad\text{and}\qquad}

\newcommand{\beq}{\begin{eqnarray}}
\newcommand{\eeq}{\end{eqnarray}}
\newcommand{\be}{\begin{eqnarray*}}
\newcommand{\ee}{\end{eqnarray*}}
\newcommand{\bal}{\begin{align}}
\newcommand{\eal}{\end{align}}
\newcommand{\rmd}{{\rm d}}

\def\mc{\text{mc}}

\def\I{\text{I}}
\def\II{\text{II}}

\DeclareMathOperator*{\argmax}{arg\,max}
\DeclareMathOperator*{\argmin}{arg\,min}

\title{High-Dimensional Unfolding in Large Backgrounds}

\author[a]{Alexandre Falc\~ao}
\author[b]{and Adam Takacs}

\affiliation[a]{Department of Physics and Technology, University of Bergen, Allegaten 55, 5007 Bergen, Norway}
\affiliation[b]{Institute for Theoretical Physics, Heidelberg University, Philosophenweg 16, 69120 Heidelberg, Germany}

\emailAdd{alexandre.falcao@uib.no}
\emailAdd{takacs@thphys.uni-heidelberg.de}

\abstract{
We propose new methodologies in multi-dimensional unfolding in dense environments, and show that incorporating auxiliary observables can significantly improve performance.
Our approach builds on the ML-based OmniFold algorithm, which we extend to account for background, detector acceptance, efficiency, and uncertainties, enabling its application in high-luminosity and heavy-ion collision settings. We derive this algorithm and demonstrate its mathematical and numerical equivalence to expectation-maximization and Iterative Bayesian Unfolding (IBU).
We illustrate our method with a realistic jet substructure analysis incorporating both large background and detector simulation. Our analysis includes up to 18 observables, leading to significantly improved performance in the unfolding. 
We propose a method that integrates calibration and unfolding into a single, consistent framework, and demonstrate enhanced performance relative to traditional methods.
These developments lay the groundwork for robust, high-dimensional, ML-based unfolding and calibration in complex collider environments across a wide range of analyses.
}

\begin{document} 
    \maketitle
\flushbottom

\section{Introduction}
\label{sec:Introduction}

Jet measurements at the LHC and RHIC are central to advancing our understanding of the strong interaction. They probe the quark-gluon plasma (QGP), enable precision tests of the Standard Model, and provide a foundation for searches for new physics~\cite{Salam:2010nqg,Larkoski:2017jix,Marzani:2019hun,Kogler:2018hem,dEnterria:2009xfs,Connors:2017ptx,Cunqueiro:2021wls}. However, these measurements rely on detector-distorted signals, requiring corrections through calibration and unfolding techniques. The dominant source of reported systematic uncertainties arises from these corrections, especially in environments with large and fluctuating backgrounds.

In proton-proton ($pp$) collisions, the background arises from pileup (multiple simultaneous interactions) which is expected to reach an average of 140 collisions per bunch crossing in the high-luminosity phase of the LHC~\cite{ZurbanoFernandez:2020cco}. In heavy-ion ($AA$) collisions, the background originates from the underlying event, whose magnitude can be equivalent to several hundred pileup collisions occurring at a single vertex.
Both pileup and the underlying event contaminate reconstructed jets, distorting their energies and producing fake jet-like signals. Subtraction techniques are used to mitigate these effects~\cite{Cacciari:2007fd,Berta:2014eza,Cacciari:2014gra,Bertolini:2014bba,STAR:2017hhs,ArjonaMartinez:2018eah,Berta:2019hnj} (see review~\cite{Soyez:2018opl}). Yet, due to the nonlinear nature of the detector response, subtraction alone is insufficient. Additional calibration and unfolding procedures are required~\cite{CMS:2011shu,CMS:2016lmd,ATLAS:2024ogz,ALICE:2023waz,ATLAS:2024jtu}. 

Unfolding in heavy-ion collisions is especially challenging. The underlying event not only introduces background, but also carries valuable information about the QGP, which modifies jet substructure and induces energy loss~\cite{Mehtar-Tani:2013pia,Blaizot:2015lma,Cao:2024pxc}. At the same time, jets deposit energy into the QGP, producing back-reaction effects~\cite{Cao:2020wlm}. While removing more of the underlying event can improve measurement precision, it risks erasing key information about the medium.

In this work, we address the challenges of unfolding in the presence of large backgrounds. Our goal is to show that increasing the dimensionality of the unfolding will improve performance, reducing systematic uncertainties in jet analyses. Our method builds on recent developments in machine learning (ML)-based unfolding~\cite{Andreassen:2019cjw,Andreassen:2021zzk,Bellagente:2020piv,Backes:2022sph,Backes:2023ixi,Diefenbacher:2023wec} (see also ML reviews~\cite{Zhou:2023pti,Larkoski:2024uoc,HEPML-LivingReview}), as they enable high-dimensional unfolding. 

We build on OmniFold~\cite{Andreassen:2019cjw}, an unbinned ML-classifier-based unfolding algorithm that recently has been applied to experimental data~\cite{H1:2021wkz,LHCb:2022rky,H1:2023fzk,Song:2023sxb,Pani:2024mgy,ATLAS:2024xxl,ATLAS:2025qtv,CMS:2025sws}. 
We present \textbf{OmniFold-HI}, an improved version of the original algorithm, tailored to account for large backgrounds, fake signals, detector acceptance, efficiency losses, and uncertainties. The inclusion of these effects is crucial for precise heavy-ion, as well as high-luminosity, analyses where these corrections are not negligible. Our derived formulas, their implementation and application to realistic data go beyond the previous extension of OmniFold~\cite{Andreassen:2021zzk}.
We provide a rigorous derivation showing that OmniFold-HI corresponds to the expectation-maximization algorithm and is equivalent to iterative Bayesian unfolding (IBU)~\cite{DAGOSTINI1995487,2010arXiv1010.0632D}. Our improvements provide a solid theoretical ground for further extensions to OmniFold. We show, step by step, that neural networks in OmniFold are computationally efficient alternatives to traditional histogram ratios, and provide direct physical meaning to the training loss functions.

Using OmniFold-HI, we perform multidimensional unfolding of jet observables in a heavy-ion background. To increase the dimensionality, we include additional observables alongside the target ones, up to 18 dimensions. We introduce a strategy to optimally select auxiliary observables by analyzing their correlation strengths. We consider several Monte Carlo (MC) event generators, detector effects, jet calibration, and background subtraction. 

We demonstrate that the use of auxiliary observables significantly improves unfolding performance compared to traditional methods. Furthermore, this approach enhances full-event unfolding, which we demonstrate by successfully unfolding observables that were not included in the simulation. This further demonstrates the robustness of this high-dimensional unfolding framework. Moreover, we compare several calibration strategies and show that integrating unfolding and jet calibration within a unified, high-dimensional framework improves performance, reducing analysis uncertainties.

All generated events, detector simulations, and analysis code are publicly available, supporting future developments in unfolding techniques and machine learning applications for heavy-ion collisions~\cite{OmniFoldHI_GitHub}.

\section{Unfolding}
\label{sec:Unfolding}

In jet analyses, the reconstructed four-momentum of jets is first calibrated to match the true value, thereby improving event selection~\cite{CMS:2017yfk,ATLAS:2020cli}. In high-background environments, this calibration involves subtracting pileup and background contributions and correcting for detector distortions using information from the reconstructed event. These corrections rely on detector simulations of MC events and are further refined using model-independent methods, such as exploiting the known momentum balance in $Z$+jet events. Once the jets are corrected, distributions of the targeted observables (e.g. substructure or correlations) are constructed. Unfolding is then applied to correct for detector effects in these target observables using MC samples and detector simulations. Both calibration and unfolding are inference tasks: the former is performed event-by-event using parameterized formulas, while unfolding is traditionally applied to distributions.

Unfolding aims to infer the probabilistic relationship between measured \textit{effects} and underlying true \textit{causes}. These causes and effects correspond to finite regions of event space, typically defined over a few observables. In the most common case, observables are binned, and probabilities are represented as histograms, referred to as binned unfolding. More recently, advanced tools have enabled more abstract representations of the event space~\cite{Komiske:2019fks,Komiske:2020qhg,Larkoski:2023qnv} and have facilitated the development of unbinned and event-by-event unfolding~\cite{Andreassen:2019cjw,Bellagente:2020piv,Backes:2023ixi,Diefenbacher:2023wec}. 

\begin{figure}
    \centering
    \includegraphics[width=\textwidth,page=1]{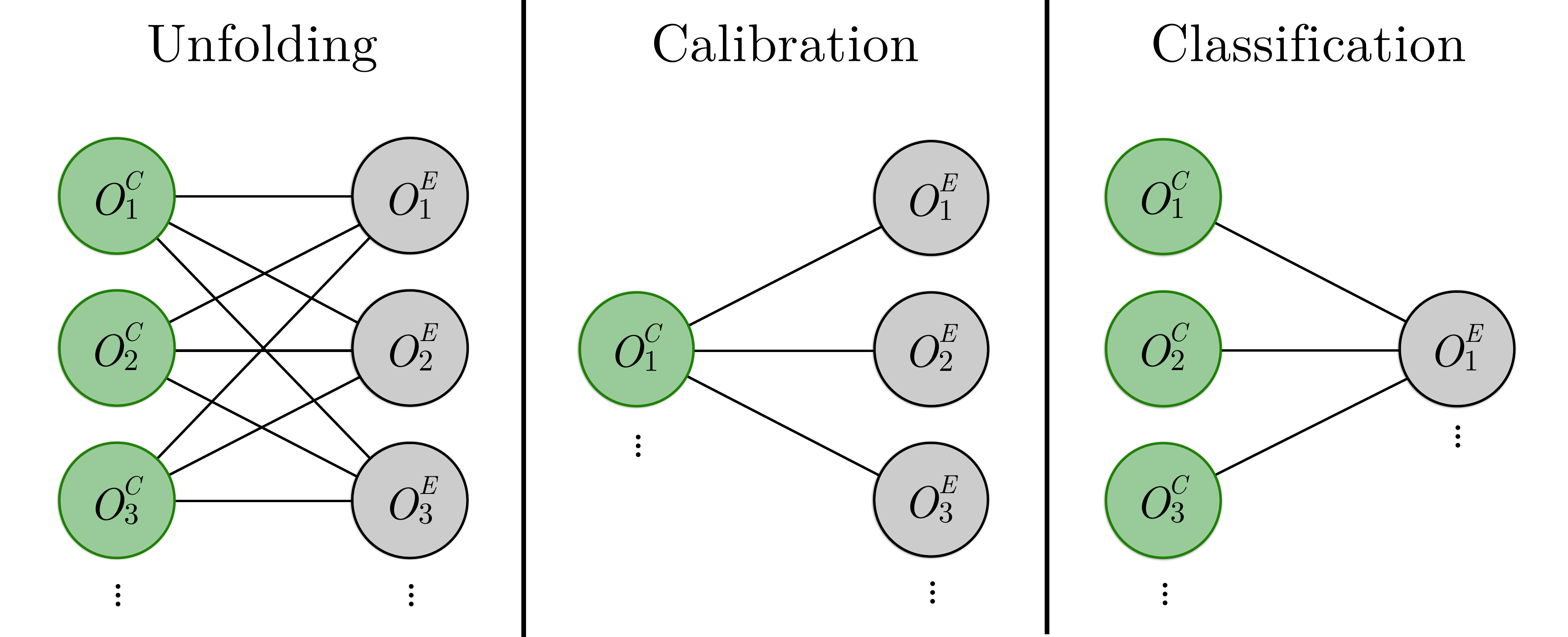}
    \caption{Unfolding, calibration, and classification as inference tasks. Circles represent observables, labels, or other quantities, with green indicating true (cause-level) values and gray indicating measured (effect-level) values. Links represent probabilistic relationships.}
    \label{fig:unfolding_sketch}
\end{figure}

Causes and effects can be multidimensional, with each dimension corresponding to an observable being unfolded (see the left panel of Fig.~\ref{fig:unfolding_sketch}). Since detector effects often introduce correlations between observables, incorporating multiple observables into the unfolding process naturally enhances both accuracy and robustness. This motivates our focus on high-dimensional unfolding, which we will demonstrate in this work. However, the computational cost of binning scales poorly with dimensionality, making high-dimensional binned unfolding impractical. In contrast, ML-based unfolding methods offer efficient scalability to higher dimensions at a significantly lower computational cost.

More generally, causes and effects can differ in dimensionality. A straightforward example of inferring multiple measured quantities into a single observable is jet calibration (i.e. jet energy scale correction), where jet momentum is inferred from tracks, calorimeter signals, and pileup properties (see the middle of Fig.~\ref{fig:unfolding_sketch}). In this case, the inference is parametrized and applied on an event-by-event basis. Conversely, mapping a measurement to multiple possible labels arises in classification problems, where a jet may originate from quarks, gluons, or heavy decays~\cite{Larkoski:2019nwj,Dreyer:2018nbf,Dreyer:2021hhr,Brewer:2018dfs,Takacs:2021bpv}, or may be quenched or unquenched~\cite{Brewer:2018dfs,Du:2020pmp,Du:2021pqa,Lai:2021ckt,CrispimRomao:2023ssj}. In these cases, the cause space contains additional labels (see the right of Fig.~\ref{fig:unfolding_sketch}). Therefore, unfolding, calibration, and classification are all multi-dimensional inference tasks that can be addressed within a unified framework. In our work, we show that combining calibration and unfolding into a single step can further enhance the performance of jet analyses. This is due to the consistent treatment of multi-dimensional correlations.

In the following, we investigate the probabilistic description of unfolding. We study the relation between causes and effects, that is usually provided by some model. 
We review the theory and our implementation of the most commonly used Iterative Bayesian Unfolding (IBU)~\cite{DAGOSTINI1995487,2010arXiv1010.0632D}, and the ML-based and unbinned OmniFold algorithm~\cite{Andreassen:2019cjw}. We show that these methods are equivalent, and are maximum likelihood estimators, as well as applications of the so-called expectation maximization algorithm. While some of these observations were previously shown, here we provide a proof and a complete derivation of both algorithms. This was necessary to introduce improvements to both algorithms to handle large background, detector acceptance, efficiency, and uncertainties.

\subsection{Iterative Bayesian unfolding (IBU)}
\label{sec:IBU}
Let us start with binned distributions, and infer the number of counts $x(C_i)$ in the $C_i$ cause-bin (truth) from the knowledge of effect (measured) counts $x(E_j)$,
\begin{equation}
    \label{eq:Unfolding_basics}
    x(C_i)=\sum_jp(C_i|E_j)\cdot x(E_j)\,,
\end{equation}
where $p(C_i|E_j)=p(C_i,E_j)/p(E_j)$, and $i,j$ denotes different bins in the space of chosen observables. Fake $F$ and Trash $T$ bins are introduced for events which don't produce cause-, and effect-counts, respectively. The probability is defined as $p(\__i)=x(\__i)/\sum_{i'} x(\__{i'})$, and Eq.~\eqref{eq:Unfolding_basics} leads to $\sum_i x(C_i)=\sum_j x(E_j)$. As trash is not measured, the efficiency factor is introduced
\begin{equation}
    \begin{split}
        x(C_i)&=\sum_{j\neq T}p(C_i|E_j)\cdot x(E_j) + p(C_i|T)\cdot x(T)\\
        &=\sum_{j\neq T}p(C_i|E_j)\cdot x(E_j) + p(T|C_i)\cdot x(C_i)\\
        &=\frac{1}{\epsilon_i}\sum_{j\neq T}p(C_i|E_j)\cdot x(E_j)\,,
    \end{split}
\end{equation}
where Bayes' identity is used in the second step, and $\epsilon_i=1-p(T|C_i)$. The conditional probability is inverted using Bayes' theorem
\begin{equation}
    \label{eq:Bayes_theorem}
    p(C_i|E_j)\approx\frac{p(E_j|C_i)\cdot p_0(C_i)}{\sum_{i'} p(E_j|C_{i'})\cdot p_0(C_{i'})}\,,
\end{equation}
where $p_0(C_i)$ denotes the prior and so $p(C_i)=\sum_j p(C_i,E_j)\neq p_0(C_i)$. For $p_0(C_i)\to p(C_i)$, the approximation becomes equality in Eq.~\eqref{eq:Bayes_theorem}. Unfolding assumes a known response matrix that is typically constructed from simulations $p(E_j|C_i)\approx p(E_j|C_i,{\rm mc})$ or, in other words, we unfold the effects described by the simulation. By evaluating Eqs.~\eqref{eq:Unfolding_basics}--\eqref{eq:Bayes_theorem} for a given prior $p_0(C_i)$, one can estimate $x(C_i)$. This algorithm clearly resembles Bayesian inference. 

To reduce the bias introduced by the prior $p_0(C_i)$, the following iterative algorithm was introduced and referred to as iterative Bayesian unfolding (IBU)~\cite{DAGOSTINI1995487,2010arXiv1010.0632D},
\begin{equation}
    \label{eq:IBU_recursion}
    x_n(C_i)=\frac{1}{\epsilon_i}\sum_{j\neq T}\frac{p(E_j|C_i,{\rm mc})\cdot p_{n-1}(C_i)}{\sum_{i'} p(E_j|C_{i'},{\rm mc})\cdot p_{n-1}(C_{i'})}\cdot x(E_j)\,,
\end{equation}
where $n$ is the number of iteration. For a single step ($n=1$), Eq.~\eqref{eq:IBU_recursion} reduces to ordinary Bayesian inference, with $p_0(C_i)$ being the prior and $x_1(C_i)$ the marginal posterior.

Although Eq.~\eqref{eq:IBU_recursion} has a Bayesian form, it is a maximum likelihood strategy for $n\to\infty$, for which the limit becomes prior independent. To show this, we identify the log-likelihood as
\begin{equation}
    \begin{split}
        \log{\rm L}[p(C_i)]&=\log\left[\sum_ip(E_j|C_i)\cdot p(C_{i})\right]\,.
    \end{split}
\end{equation}
Averaging the expected log-likelihood over possible measurements results, and including probability conservation with a Lagrange multiplier (see also App.~A in~\cite{Andreassen:2019cjw})
\begin{equation}
    \begin{split}
        {\rm LL}[p(C_i)]&=\sum_jp(E_j)\log\left[\sum_ip(E_j|C_i)\cdot p(C_{i})\right]-\lambda\left(\sum_{i}p(C_{i})-1\right)\,.
    \end{split}
\end{equation}
The maximum of this likelihood results in a cause probability of which effect is the closest to the measurement. The maximum corresponds to $\frac{\delta\log{\rm LL}}{\delta p(C_i)}=0$, resulting in $\lambda=1$, and
\begin{equation}\label{eq:pstar}
    p^*(C_i)=\sum_j\frac{p(E_j|C_i)\cdot p^*(C_i)}{\sum_{i'}p(E_j|C_{i'})\cdot p^*(C_{i'})}\cdot p(E_j)\,.
\end{equation} 
The solution is indeed the maximum as $\frac{\delta^2{\rm LL}}{\delta p(C_{i_1})\cdot\delta p(C_{i_2})}<0$. The same maximum can be derived by maximizing any rescaled version of the likelihood (e.g. the cross-entropy~\cite{Nachman:2021yvi}). In what follows, we show that Eq.~\eqref{eq:IBU_recursion} recovers Eq.~\eqref{eq:pstar} for $n\to \infty$, corresponding to the expectation-maximization (EM) algorithm that is a known strategy to maximize the likelihood. 

Our goal is to show that Eq.~\eqref{eq:IBU_recursion} increases the likelihood (${\rm LL}[p_{n+1}]\geq{\rm LL}[p_n]$) towards its maximum, Eq.~\eqref{eq:pstar}. We introduce the auxiliary distribution $q(C_i|E_j)$,
\begin{equation}
    \begin{split}
        \log\left(\sum_ip(E_j|C_i)\cdot p(C_i)\right) &= \log\left(\sum_iq(C_i|E_j)\frac{p(E_j|C_i)\cdot p(C_i)}{q(C_i|E_j)}\right) \\
        &\geq \sum_iq(C_i|E_j)\cdot \log\frac{p(E_j|C_i)\cdot p(C_i)}{q(C_i|E_j)}\,.
    \end{split}
\end{equation}
In the last line we used Jensen's inequality. The likelihood is therefore bounded from below for $q$. To measure the distance between $q$ and $p$, we introduce the cross-entropy-like object
\begin{equation} 
    Q[p(C_i),q(C_i|E_j)]=\sum_j p(E_j)\sum_i q(C_i|E_j)\log \frac{p(E_j|C_i)p(C_i)}{q(C_i|E_j)}\,,
\end{equation}
for which we know ${\rm LL}[p]\geq Q[p,q]$, and the equality comes from $q\to p$. Following the EM algorithm, we estimate $q$ from Eq.~\eqref{eq:IBU_recursion} (E-step),
\begin{equation}
    q(C_i|E_j) = \frac{p(E_j|C_i)\cdot p_{n}(C_i)}{\sum_{i'} p(E_j|C_{i'})\cdot p_{n}(C_{i'})}\,,
\end{equation}
and then maximize $Q[p,q]$ (M-step),
\begin{equation}
    p_{n+1}(C_i) = \argmax_{p(C_i)}Q[p,p_n]=\sum_j\frac{p(E_j|C_i)\cdot p_n(C_i)}{\sum_{i'}p(E_j|C_{i'})\cdot p_n(C_{i'})}\cdot p(E_j)\,.
\end{equation}
In the last step, we evaluated the maximum, arriving to the IBU recursion from Eq.~\eqref{eq:IBU_recursion}. Maximizing $Q$ therefore involves the maximization of the likelihood
\begin{equation}
    {\rm LL}[p_n]=Q[p_n,p_n]\leq Q[p_{n+1},p_n]\leq{\rm LL}[p_{n+1}]\,.
\end{equation}
The iterative Bayesian unfolding is therefore equivalent to the EM algorithm. This also explains why $p_n$ becomes independent of the prior $p_0$ as $n\to\infty$, which could be surprising for the seemingly Bayesian inference. 

During unfolding, $p(E_j|C_i,{\rm mc})$ is evaluated as a multidimensional histogram using MC events, background embedding and detector simulation. Generated events also usually serve as prior $p_0(C_i)$. Evaluating Eq.~\eqref{eq:IBU_recursion} is a simple linear algebra task, while the computational bottleneck of multi-dimensional IBU is constructing the histograms. Trash and Fakes include kinematic cuts during event generation, and the possibility of under/overflowing bins. Moreover, bins must be filled meaningfully to predict the probability density, and therefore, an increasing number of simulated events is needed. Further complications with IBU arise if the response matrix is not uniquely invertible.\footnote{If present, non-invertibility is inherent to the inference task and thus poses a challenge not only for IBU but for any unfolding algorithm.} Understanding the effect of finite iterations, imperfect maximum finding, and competing statistical uncertainties is also exciting. We study these complications in the next section on our unfolding example.

We tested our implementation up to 3-4 observables with 10-20 bins and 1M events, and the bottleneck was the histogram function in \texttt{numpy}, which eventually ran out of memory. One can access more computing power or use more efficient histograms, and thus unfolding in a slightly higher dimensions with IBU is accessible, although still greatly limited by the available memory. In what follows, we turn to neural networks to generalize the IBU algorithm to much higher dimensions.

\subsection{The OmniFold-HI algorithm}
\label{sec:OmniFold}

We extend IBU to the entire event phase space by inferring the number density of events
\begin{equation}
    x(t) = \int\rmd m \, p(t|m)\,x(m)\,,
\end{equation}
where $t$ and $m$ denotes truth and measured event phase space (previously cause and effect). The number of events is fixed $\int\rmd m\,x(m)\equiv\int\rmd t\, x(t)=N_{\rm ev}$, and $p(t|m)=p(t,m)/p(m)$, where the probabilities are $p(\_)=x(\_)/N_{\rm ev}$. To recover Sec.~\ref{sec:IBU}, events have to be binned, $x(C_i) = \int\rmd t \, x(t\in C_i)$, $x(E_j) = \int\rmd m \, x(m\in E_j)$. Following Eq.~\eqref{eq:IBU_recursion}, the generalized version of the IBU algorithm is
\begin{equation}
    \label{eq:of_recursion}
    x_n(t) = \frac{1}{\epsilon(t)}\int_{m\neq T}\rmd m \, \frac{p(m|t)\,x_{n-1}(t)}{\int\rmd t'\,p(m|t')\,x_{n-1}(t)}\,x(m)\,.
\end{equation}
Equivalently to IBU, the efficiency factor $\epsilon(t)=1-p(m=T|t)$ compensates for unmeasured (trash) events. The total number of events is conserved, $\int\rmd t\, x_n(t)=N_{\rm ev}$. We introduce the prediction $x_n(m)=\int\rmd t\,p(m|t)\,x_n(t)$.

In Sec.~\ref{sec:IBU}, we showed Eq.~\eqref{eq:of_recursion} is Bayesian inference for $n=1$ with $x_0(t)$ prior and $x_1(t)$ marginal posterior. For $n\to\infty$, $x_n(t)$ maximizes the likelihood and the iteration corresponds to the EM-algorithm. The impracticality of Eq.~\eqref{eq:of_recursion} lies in the construction densities. In contrast to the binned case, $t$ and $m$ cover a high-dimensional space that is inefficient to sample. The estimated size of the event phase space is $\sim400$ dimensions for 100 reconstructed particles based on their 4-momenta. In a simpler setting, the event space is recused to its summary statistics, lets say to 20 observables. Then, $t$ and $m$ are 20-dimensional arrays. In what follows, we show that, with the usage of neural networks and machine learning, one can evaluate Eq.~\eqref{eq:of_recursion} efficiently.

Instead of densities, let us rewrite Eq.~\eqref{eq:of_recursion} using the likelihood ratios $\omega_n(m)=x(m)/x_{n-1}(m)$ and $\nu_n(t)=x_{n}(t)/x_0(t)$, 
\begin{equation}
    \begin{split}
        x_n(t)& = \frac{1}{\epsilon(t)}\int_{m\neq T}\rmd m \, p(m|t)\,\nu_{n-1}(t)\,\omega_n(m)\cdot x_0(t)\,,
    \end{split}
\end{equation}
and a similar equation for $x_n(m)$. Eq.~\eqref{eq:of_recursion} translates to the following two step recursion
\begin{equation}
    \label{eq:of_algorithm}
    \begin{split}
        {\rm Step\,\, I}:&\hspace{1cm} \omega_n(m)=\frac{x(m)}{x_{n-1}(m)}=\frac{x(m)}{\nu^\text{push}_{n-1}(m)\cdot x_0(m)}\,,\\
        &\hspace{1cm} \omega^\text{pull}_n(t) = \int_{m\neq T}\rmd m\, p(m|t)\,\omega_n(m)\,, \\
        {\rm Step\,\, II}:&\hspace{1cm} \nu_n(t)=\frac{\omega_n^\text{pull}(t)\cdot x_{n-1}(t)}{\epsilon(t)\cdot x_0(t)}=\frac{\omega_n^\text{pull}(t)\,\nu_{n-1}(t)\cdot x_0(t)}{\epsilon(t)\cdot x_0(t)}\,, \\
        &\hspace{1cm} \nu^\text{push}_n(m) = \int\rmd t\, p(t|m)\,\nu_n(t)\,, \\
        {\rm then}:&\hspace{1cm}
        \begin{cases}
                x_n(t)=\nu_{n}(t) \cdot x_0(t)\,,\\
                x_n(m)=\nu^\text{push}_n(m) \cdot x_0(m)\,.
        \end{cases}
    \end{split}
\end{equation}
In the first step $\nu_0^{\rm push}(m)=1$, then $\omega_1(m)$ and $\omega^{\rm pull}_1(t)$ are evaluated. Afterwards, $\nu_1(t)$ and $\nu_1^{\rm push}(m)$ are evaluated and the recursion is restarted with the updated Step I. The construction of likelihood ratios in $\omega_n$ and $\nu_n$ are computationally expensive and we turn to alternative methods.

The OmniFold algorithm~\cite{Andreassen:2019cjw} implements a simplified version of Eq.~\eqref{eq:of_algorithm} using neural network classifiers, as those evaluate likelihood ratios computationally effectively as we will show.\footnote{
    The appendix of Ref.~\cite{Andreassen:2019cjw} already establishes the equivalence between IBU and OmniFold. Building upon this, we present a related derivation that explicitly incorporates Fake and Trash events, which are essential for accurately modeling acceptance and efficiency. While some aspects of this extended formulation were previously introduced in Ref.~\cite{Andreassen:2021zzk}, mathematical proofs were omitted. In contrast, our work provides a complete derivation of the algorithm, clarifies the roles of the neural network components, and outlines strategies for handling uncertainties. As such, our results extend and complement the earlier developments.} 
The likelihood ratios $\omega_n(m)$ and $\nu_n(t)$ can be obtained by introducing the functions $c_\I(m)$ and $c_\II(t)$ and minimizing the following loss functionals\footnote{
    Here we use binary-cross-entropy but others losses work as well~\cite{Nachman:2021yvi,Rizvi:2023mws}.}
\begin{equation}
\label{eq:loss}
    \begin{split}
        & \mathcal L_\I[c_\I(m)] = -\int\rmd m \left[ x(m)\cdot \log(c_\I(m)) + x_{n-1}(m)\cdot \log(1-c_\I(m)) \right]\,,\\
        & \mathcal L_\II[c_\II(t)] = -\int\rmd t \left[ \omega^\text{pull}_n(t)\,x_{n-1}(t)\cdot \log(c_\II(t)) + \epsilon(t)\,x_0(t)\cdot \log(1-c_\II(t)) \right]\,,
    \end{split}
\end{equation}
where $0\leq c_{\I,\II}\leq 1$. The minimum, $c^*(m)=\argmin_{c(m)}\mathcal L[c(m)]$, satisfies 
\begin{equation}
    \begin{split}
        & \omega_n(m) = \frac{c^*_\I(m)}{1-c^*_\I(m)}\,,\\
        & \nu_n(t) = \frac{c^*_\II(t)}{1-c^*_\II(t)}\,.
    \end{split}
\end{equation}
where we used $\delta \mathcal L/\delta c=0$. Therefore, instead of evaluating the likelihood ratios in Eq.~\eqref{eq:of_algorithm}, one could minimize the loss functionals of Eq.~\eqref{eq:loss}. 

In practice, we can only sample the event phase space. The samples correspond to measured events and their weights $(m_i,w_i)_{\rm meas}$, and so $x(m)=\sum_{i\in{\rm meas}}w_i\,\delta(m-m_i)$ and $N_{\rm ev}=\sum_{i\in{\rm meas}}w_i$. Monte Carlo generated events and their detector simulation are also accessible $(t_i,m_i,w_i)_{\rm mc}$, and $N_{\rm ev}=\sum_{i\in\mc}w_i$.\footnote{
    In the case of $N_{\rm ev}\neq N_{\rm mc}$, we redefine MC weights such that $N_{\rm ev}=\sum_{i\in\mc}w_i$. This correction can be estimated using the prior as $N_{\rm ev}\approx N_{\rm mc}N_{\rm meas\neq T}/N_{\rm mc\neq T}$. In practice, Step I of the first iteration automatically sets the normalization to $N_{{\rm ev}\neq T}$. 
}
One has to note, MC events are passed through simulation one-by-one, so $p(m^\mc_i|t^\mc_j)=\delta_{ij}$. Eq.~\eqref{eq:loss} for samples becomes
\begin{equation}\label{eq:loss_finite_1}
    \begin{split}
        \mathcal L_\I[c_\I(m)] &= -\sum_{i}^{\rm meas} w_i\cdot \log(c_\I(m_i)) - \sum_{i}^{\mc} w_i\,\nu_{n-1}^{\rm push}(m_i)\cdot \log(1-c_\I(m_i))\,,\\
        \mathcal L_\II[c_\II(t)] &= -\sum_{i}^{\mc} w_i\,\omega^\text{pull}_n(t_i)\,\nu_{n-1}(t_i)\cdot \log(c_\II(t_i)) -\sum_{i}^{\mc}w_i\,\epsilon(t_i)\cdot \log(1-c_\II(t_i))\,.
    \end{split}
\end{equation}
As the measurement has no information of Trash, both measured and MC events for which $m_i=T$ can be skipped Step I minimization, and $\omega_n(m_i=T)=0$ is set. Similar simplification can be done in Step II minimization, by removing all MC truth for which $m_i=T$. Pulling and pushing weights on the MC level is trivial $\omega_n^{\rm pull}(t_i^\mc)=\omega_n(m_i^\mc)$ and $\nu_n^{\rm push}(m_i^\mc)=\nu_n(t_i^\mc)$. The efficiency factor becomes simply $\epsilon(t_i^\mc)=\delta_{m_i^\mc\neq T}$. As a reminder, here Trash and Fakes are from kinematic cuts during event generation, and not only from over/underflow bins. Going from Eq.~\eqref{eq:loss} to \eqref{eq:loss_finite_1} corresponds to the Monte Carlo evaluation of the loss integrals.

To implement Eq.~\eqref{eq:loss_finite_1}, one has to parametrize $c_\I(m)$ and $c_\II(t)$ that span over the huge event phase space and tune its parameters to minimize the loss functions. We use deep neural networks as they can approximate most functions (see universal approximation theory). Furthermore, Eq.~\eqref{eq:loss_finite_1} can be rewritten as the loss function of weighted binary classifiers:
\begin{equation}\label{eq:loss_finite_2}
    \begin{split}
        \mathcal L_\I[c_\I(m)] &= -\sum_{i\neq T}^{\rm meas+mc} w^\I_i \cdot\Big[ \delta_{i\in{\rm meas}}\cdot \log(c_\I(m_i))+\delta_{i\in{\rm mc}}\cdot \log(1-c_\I(m_i)) \Big]\,,\\
        \mathcal L_\II[c_\II(t)] &= -\sum_{i\neq T}^{\rm mc+mc} w^\II_i\cdot \Big[ \delta_{i\in\mc_1}\cdot \log(c_\II(t_i))+\delta_{i\in\mc_2}\cdot \log(1-c_\II(t_i)) \Big]\,,
    \end{split}
\end{equation}
where $\delta_{i\in{\rm \ell abel}}=0,1$ acts as labels, and the weights are
\begin{equation}
    \begin{split}
        w_i^\I&=\delta_{i\in{\rm meas}}\,w_i^{\rm meas} + \delta_{i\in\mc}\,w_i^\mc\nu_{n-1}^{\rm push}(m_i^\mc)\,,\\
        w_i^\II&=\delta_{i\in\mc_1}\,w_i^\mc\,\omega_n^{\rm pull}(t^\mc_i)\,\nu_{n-1}(t^\mc_i)+\delta_{i\in\mc_2}\,w_i^\mc\,\epsilon(t^\mc_i)\,.
    \end{split}
\end{equation}
The minimization of Eq.~\eqref{eq:loss_finite_2} corresponds to training two DNN classifiers to distinguish between measured and MC events in step I, and MC and MC events in step II, with their appropriate weights. Here $\delta_{i\in{\rm \ell abel}}=0,1$ are the known input labels, and $c_\I(m)$ and $c_\II(t)$ are the predictions. Using appropriate loss functions turns NN training into evaluating histogram ratios.

Our final \texttt{OmniFold-HI} algorithm is summarized as
\begin{equation}
    \label{eq:OFHI_algorithm}
    \begin{split}
        {\rm Step\,\, I}:&\hspace{1cm} \omega_n(m)=\frac{c^*_\I(m)}{1-c^*_\I(m)}, \hspace{0.5cm}\text{where}\hspace{0.5cm} c^*_\I(m)=\argmin_{c_\I(m)} \mathcal L_\I\,,\\
        &\hspace{1cm} \omega_n^\text{pull}(t^\mc_i)=\delta_{m^\mc_i\neq T}\,\omega_n(m^\mc_i)\,, \\
        {\rm Step\,\, II}:&\hspace{1cm} \nu_n(t)=\frac{c^*_\II(t)}{1-c^*_\II(t)},\hspace{0.5cm}\text{where}\hspace{0.5cm}c^*_\II(t)=\argmin_{c_\II(t)} \mathcal L_\II\,, \\
        &\hspace{1cm} \nu^\text{push}_n(m^\mc_i)=\nu_n(t^\mc_i)\,, \\
        {\rm then}: &\hspace{1cm} x_n(t)=\sum_i^{\rm mc}w_i\,\nu_{n}(t_i)\,\delta(t-t_i)\,,\\
        &\hspace{1cm} x_n(m)=\sum_i^{\rm mc}w_i\,\nu^\text{push}_n(m_i)\,\delta(m-m_i)\,.
    \end{split}
\end{equation}
We would like to emphasize Eq.~\eqref{eq:OFHI_algorithm} is the unbinned equivalent of IBU from Eq.~\eqref{eq:IBU_recursion}. The resulting reweighted samples then can be used to obtain all sorts of quantities, naturally predicting Fake, Trash probabilities, and $N_{\rm ev}$. After binning, acceptance and bin migration (under/overflow) are also available. Reference~\cite{Andreassen:2021zzk} has already provided a strategy for OmniFold to handle Fakes, Trash, however, it lacks derivations, and introduces additional neural networks.\footnote{
    Reference~\cite{Andreassen:2021zzk} mentions the possibility of avoiding additional NNs by using the prior, similar to our method.
} Our new algorithm avoids additional NNs, relies on mathematical formulas, and provides a robust framework for future developments.

\subsection{Unfolding uncertainties}
\label{sec:Uncertainties}

Unfolding imperfections are a common source of systematic uncertainties in jet analyses. These uncertainties can be broadly categorized into those arising from the propagation of statistical fluctuations and those due to imperfect modeling. In this section, we review these sources of uncertainty and discuss strategies for quantifying them. As we will demonstrate, traditional methods of uncertainty estimation must be applied with care when using OmniFold, motivating the development of new approaches that we present here.

\subsubsection{Statistical uncertainties}
\label{sec:StatUnc}
Statistical uncertainties affect the measurement $x(E_j)$, the response matrix $p(E_j|C_i)$, and the prior $p_0(C_i)$. Focusing first on the measurement, it is standard to assume Poisson distributed (and independent) bin counts, such that $\Delta x(E_j)=\sqrt{x(E_j)}$.\footnote{
    For weighted events (relevant for prior and response matrix), $\Delta x(E_j)=\sqrt{\sum w^2(E_j)}$.
} A common method for propagating statistical uncertainties through unfolding is Poisson bootstrapping: to each event in the measured distribution is assigned a Poisson weight $w$ (with $\langle w\rangle=1$), and the unfolding procedure is repeated multiple times using varied realizations of $x(E_j)$. The unfolded distribution $x(C_i)$ is then taken as the average across these samples, with statistical uncertainty $\Delta x(C_i)$ estimated from the standard deviation. A faster, yet equivalent, method approximates this by directly fluctuating the bin contents of the fixed histogram $x(E_j)$ using $\Delta x(E_j)$. This approach can also be extended to include statistical fluctuations in the response matrix and the prior, and so to include uncertainty correlations. Bins shall be chosen such they contain a meaningful number of events. The resulting propagated uncertainties are typically larger than the statistical uncertainty of the underlying truth, and they increase over iterations. This is a known limitation of IBU and some sort of regularization (e.g. finite iteration or smoothing) is applied. We will see an example of this increase in Sec.~\ref{sec:Results}.

Applying the Poisson bootstrap technique to OmniFold reveals a subtle but important issue: the resulting statistical uncertainty of the unfolded result is systematically smaller than the one obtained with IBU, and smaller than the statistical uncertainty of the truth distribution itself. This behavior arises from the nature of neural network training, which is designed to suppress overfitting and thereby reduces statistical fluctuations and outliers. Further details are given in App.~\ref{sec:StatUnc_OFHI}, where we show that a more reliable uncertainty estimate is obtained using Poisson bootstrap via resampling rather than via the usual reweighing. We also demonstrate that increased NN overfitting improves the propagation of statistical fluctuations. A direct comparison with the statistical uncertainty of the truth distribution, also presented in App.~\ref{sec:StatUnc_OFHI}, indicates that the commonly used Poisson reweighting approach significantly underestimates statistical uncertainties in this setting. To the best of our knowledge, this limitation of the OmniFold has not been previously discussed.

We adopt an alternative novel strategy introduced in Ref.~\cite{Nachman:2020fff} that avoids bootstrapping altogether. Instead of propagating the uncertainty from $x(E_j)$, this method directly infers the statistical uncertainty of the truth distribution. The central idea is that, just as unfolding estimates $x(C_i)$ from $x(E_j)$, one can perform a second unfolding to infer $\Delta x(C_i)$ from $\Delta x(E_j)$, where we used Poisson bootstrap to estimate $\Delta x(C_i),\Delta x(E_j)$. Although this procedure is formally equivalent to the Poisson bootstrap, its numerical implementation mitigates NN limitations and as we will see yields more reliable uncertainty estimates. As shown in App.~\ref{sec:StatUnc_OFHI}, directly unfolding the statistical uncertainty provides a more accurate estimates. Moreover, since the unfolding algorithm needs to be executed only twice (once for the nominal and once for the uncertainty), the computational cost is significantly reduced. Appendix~\ref{sec:StatUnc_OFHI} shows little difference in the nominal unfolded results across methods, while it reveals substantial differences in the estimated uncertainties, with direct unfolding providing the most accurate truth-level statistical uncertainty.

In the case of OmniFold-HI, the resulting algorithm can run in parallel with unfolding, and it follows as 
\begin{equation}
    \label{eq:OFHI_algorithm_square}
    \begin{split}
        {\rm Step\,\, I}:&\hspace{0.5cm} \omega^2_n(m)=\frac{c^*_\I(m)}{1-c^*_\I(m)}, \hspace{0.5cm}\text{where}\hspace{0.5cm} c^*_\I(m)=\argmin_{c_\I(m)} \mathcal L^{(2)}_\I\,,\\
        &\hspace{0.5cm} \omega_n^{2,\text{pull}}(t^\mc_i)=\delta_{m^\mc_i\neq T}\,\omega^2_n(m^\mc_i)\,, \\
        {\rm Step\,\, II}:&\hspace{0.5cm} \nu^2_n(t)=\frac{c^*_\II(t)}{1-c^*_\II(t)},\hspace{0.5cm}\text{where}\hspace{0.5cm}c^*_\II(t)=\argmin_{c_\II(t)} \mathcal L^{(2)}_\II\,, \\
        &\hspace{0.5cm} \nu^{2,\text{push}}_n(m^\mc_i)=\nu^2_n(t^\mc_i)\,, \\
        {\rm then}: &\hspace{0.5cm} \Delta x_n(t)=\sqrt{x^2_n(t)}, \hspace{0.5cm}\text{with}\hspace{0.5cm}x^2_n(t)=\sum_i^{\rm mc}w_i^2\,\nu^2_{n}(t_i)\,\delta(t-t_i)\,,\\
        &\hspace{0.5cm} \Delta x_n(m)=\sqrt{x^2_n(m)}, \hspace{0.5cm}\text{with}\hspace{0.5cm}x^2_n(m)=\sum_i^{\rm mc}w_i^2\,\nu^{2,\text{push}}_n(m_i)\,\delta(m-m_i)\,.
    \end{split}
\end{equation}
The classifiers of step I and II are now trained to minimize the losses
\begin{equation}
    \label{eq:loss_finite_2_square}
    \begin{split}
        \mathcal L^{(2)}_\I[c_\I(m)] &= -\sum_{i}^{\rm meas+mc} w^{2,\I}_i \cdot\Big[ \delta_{i\in{\rm meas}}\cdot \log(c_\I(m_i))+\delta_{i\in{\rm mc}}\cdot \log(1-c_\I(m_i)) \Big]\,,\\
        \mathcal L^{(2)}_\II[c_\II(t)] &= -\sum_{i}^{\rm mc+mc} w^{2,\II}_i\cdot \Big[ \delta_{i\in\mc_1}\cdot \log(c_\II(t_i))+\delta_{i\in\mc_2}\cdot \log(1-c_\II(t_i)) \Big]\,,
    \end{split}
\end{equation}
where, as before, $\delta_{i\in{\rm \ell abel}}=0,1$ acts as labels, and the weights are non-trivial
\begin{equation}
    \begin{split}
        w_i^{2,\I}&=\delta_{i\in{\rm meas}}\,(w_i^{\rm meas})^2+\delta_{i\in\mc}\delta_{i\in\mc}\,(w_i^\mc)^2\nu_{n-1}^{2,\rm push}(m_i^\mc)\,,\\
        w_i^{2,\II}&=\delta_{i\in\mc_1}\,(w_i^\mc)^2\,\omega_n^{2,\rm pull}(t^\mc_i)\,\nu^2_{n-1}(t^\mc_i)+\delta_{i\in\mc_2}\,(w_i^\mc)^2\,\epsilon(t^\mc_i)^2\,.
    \end{split}
\end{equation}
Interestingly, when the measured and the MC weights are one by default, the uncertainty algorithm becomes equivalent with Eq.~\eqref{eq:OFHI_algorithm}, so one simply has to take the square of those weights.

The result of the statistical uncertainty unfolding can be complemented by, for example, the addition of NN training uncertainties or bootstrapping. However, it is not clear yet how to correctly piece together uncertainty propagation in OmniFold. The NN training introduces a regularization to not to be sensitive to statistical outliers. This is similar to smoothing histograms in IBU between iterations. A consistent picture to translate IBU and OmniFold-HI uncertainties is still needed and left for future work. A comparison of the bootstrapping and direct unfolding method is presented in App.~\ref{sec:StatUnc_OFHI}.

\subsubsection{Systematic uncertainties}
To quantify modeling uncertainties, analyses typically employ multiple MC simulations, varying both the response matrices and priors, and repeating the unfolding for each configuration. It is crucial that these MC samples adequately cover the relevant phase space. As more observables are considered, this requirement becomes more strict, often necessitating additional samples and careful design. For instance, jet quenching usually leads to higher jet multiplicities. Using reweighted vacuum jet samples may demand large or biased event samples to properly represent large multiplicities.

The number of iterations used in the unfolding procedure also contributes to modeling uncertainties. A further complication is that increasing the number of iterations amplifies statistical fluctuations that is a well-known feature of the IBU algorithm. This is typically addressed by limiting the number of iterations or by applying smoothing techniques between iterations to stabilize the results.

OmniFold introduces additional sources of systematic uncertainty due to the imperfect nature of NN parameterizations. These uncertainties in general can be estimated by varying the network architecture or hyperparameters and repeating the training multiple times. An alternative, more systemized approaches involve the use of Ray Tune or Bayesian neural networks for quantifying NN uncertainties~\cite{liaw2018tune,Plehn:2022ftl}.

\section{Results}
\label{sec:Results}

\subsection{Datasets}
The cause-level data (also referred to as ``truth'', or ``generated'') consists of leading-order dijet events in $pp$ collision at 5.02 TeV, generated using \texttt{Pythia8.3.12}~\cite{Skands:2014pea,Sjostrand:2014zea} and \texttt{Herwig7.3.0}~\cite{Bellm:2015jjp,Bewick:2023tfi} with default settings. To construct effect-level dataset (also called ``measured'', or ``simulated''), we mix generated events with additional particles using \texttt{JetToyHI/thermalEvent}~\cite{JetToyHI,Andrews:2018jcm} which approximates the underlying event in heavy-ion collisions.\footnote{
    Backgrounds from other sources, such as TennGen~\cite{Hughes:2020lmo} or pile-up, are deferred for future works.
} The background particles, consisting of $\pi^{\pm}$, and $\gamma$, are distributed uniformly in $\eta$ and $\phi$, and exponentially in $p_T$, emulating central $PbPb$ collisions at 5.02 ATeV using default settings.
The mixed events are processed through a CMS detector simulation using \texttt{Delphes-3.5.0}~\cite{deFavereau:2013fsa}, resulting in the effect-level dataset.

Jet reconstruction is performed in both the generated and the simulated datasets using the anti-kt algorithm with radius parameter $R=0.4$, as implemented in \texttt{fastjet-3.4.2}~\cite{Cacciari:2008gp,Cacciari:2011ma}. In \texttt{Delphes} (with the runcard form \cite{hep-lbdl/OmniFold}), jet reconstruction uses eflow objects, incorporating calorimeter and tracker efficiencies, smearing, and corrections from merging effects. We extended \texttt{Delphes} to support substructure observables from \texttt{fjcontrib-1.054}. Only the leading jet per event is retined, satisfying $p_T^{\rm jet}>250$ GeV, and $|\eta_{\rm jet}|<2.8$. The following jet observables are extracted:
\begin{itemize}
    \setlength\itemsep{0em}
    \item Jet 4-momentum observables: $p_T,\eta,\phi, m$, and multiplicity $n$.
    \item Dynamical Grooming ($\alpha=1$)~\cite{Mehtar-Tani:2019rrk,Caucal:2021bae}: $z_{g},\theta_{g},k_{t,{g}}$.
    \item Soft Drop ($\beta=0$, $z_{\rm cut}=0.1$)~\cite{Larkoski:2014wba}: $p_{T}^{sd},m_{sd},z_{sd}$.
    \item Recursive Soft Drop ($\beta=0$, $z_{\rm cut}=0.1$, $n_{\rm it}=-1$)~\cite{Dreyer:2018tjj}: $p_{T}^{rsd},n_{rsd}$.
    \item N-subjettiness ($\beta=1$, $k_t$-axis)~\cite{Thaler:2010tr}: $\tau_1,\tau_2,\tau_3,\tau_4,\tau_5$. 
\end{itemize}

Cause (effect) events where the leading jet does not pass the jet clustering $p_T$ and $\eta$ cuts are labeled as ``fake'' (``trash'') events. This results in one cause-effect jet pair per event. We did not match jets geometrically but instead unfold in $\eta,\phi$, thus demonstrating the integration of calibration and unfolding into a unified step. These event pairings also include the possibility of reconstructing a jet originating from background particles.

\begin{figure}
    \centering
    \includegraphics[height=.27\linewidth, page=1]{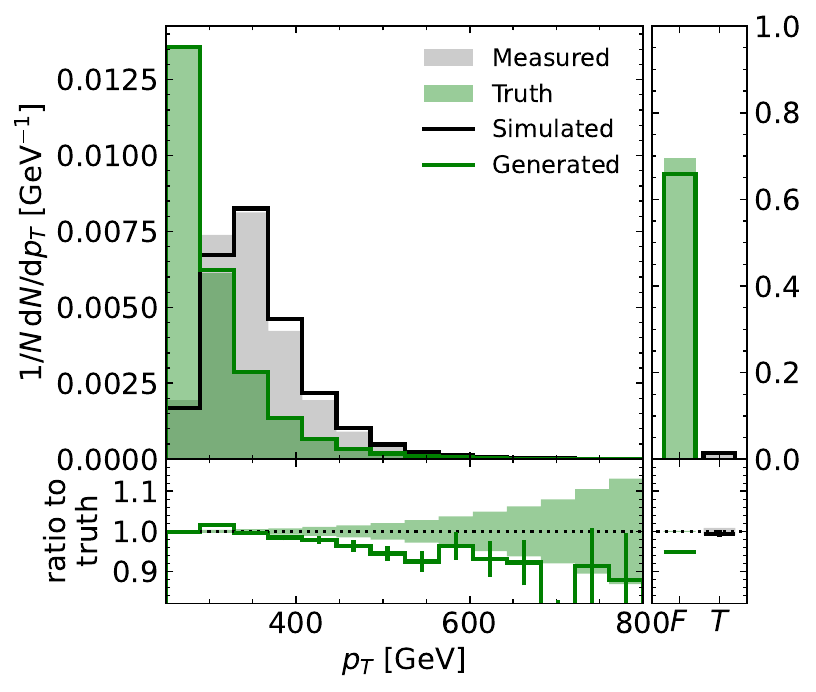}
    \includegraphics[height=.27\linewidth, page=8]{figs/data.pdf}
    \includegraphics[height=.27\linewidth, page=15]{figs/data.pdf}
    \caption{A selection of observables representing our datasets. Truth and measured is \texttt{Herwig} (+ Bkg. + \texttt{Delphes}), while generated and simulated is \texttt{Pythia} (+ Bkg. + \texttt{Delphes}). Right-hand panels show fake and trash probabilities.}
    \label{fig:data}
\end{figure}

We focus our analysis on the following three observables: the jet transverse momentum $p_T$, the Dynamical Groomed relative transverse momentum $k_{t,g}=z_g\theta_gp_T$, and the two-subjettiness $\tau_2$. The observable $k_{t,g}$, which characterizes the hardest splitting, is of current interest in both $pp$ and heavy-ion collisions due to its calculability and sensitivity to QGP structure~\cite{Caucal:2021cfb,Wang:2022yrp,Cunqueiro:2023vxl,ALICE:2024fip}. The $\tau_2$ captures the degree of two-prongness, relevant for boosted heavy object tagging~\cite{Thaler:2010tr,Larkoski:2019nwj,Marzani:2019hun}. Our selection of observables is typical of multidimensional jet substructure analyses~\cite{ATLAS:2020bbn,ALICE:2021aqk,ALICE:2022hyz,ATLAS:2022vii}.

Figure~\ref{fig:data} illustrates the three observables just described. All 18 observables can be found in the auxiliary files.
The addition of background drastically modifies the jet observables. Observables such as $p_T,k_{t,g},$ and $\tau_2$ gets shifted toward larger values, while detector effects (e.g. inefficiencies and smearing) introduce further modifications. Statistical uncertainties are small, as shown by green bands and bars in the ratio plots. In Fig.~\ref{fig:data}, we compare results from \texttt{Herwig} and \texttt{Pythia} and observe similar, though not identical, distributions (filled histograms vs lines).
For each observable, the right panels in Fig.~\ref{fig:data} show fake/trash probabilities. Notably, $\approx70\%$ of high-$p_T$ jets in the effect-level data are fake, originating from sub-threshold truth jets. This emphasizes the critical role of fakes in heavy-ion analyses. Owing to the high momentum threshold, fully combinatorial jets from the background are rare.

\begin{figure}
    \centering
    \includegraphics[width=0.32\linewidth, page=1]{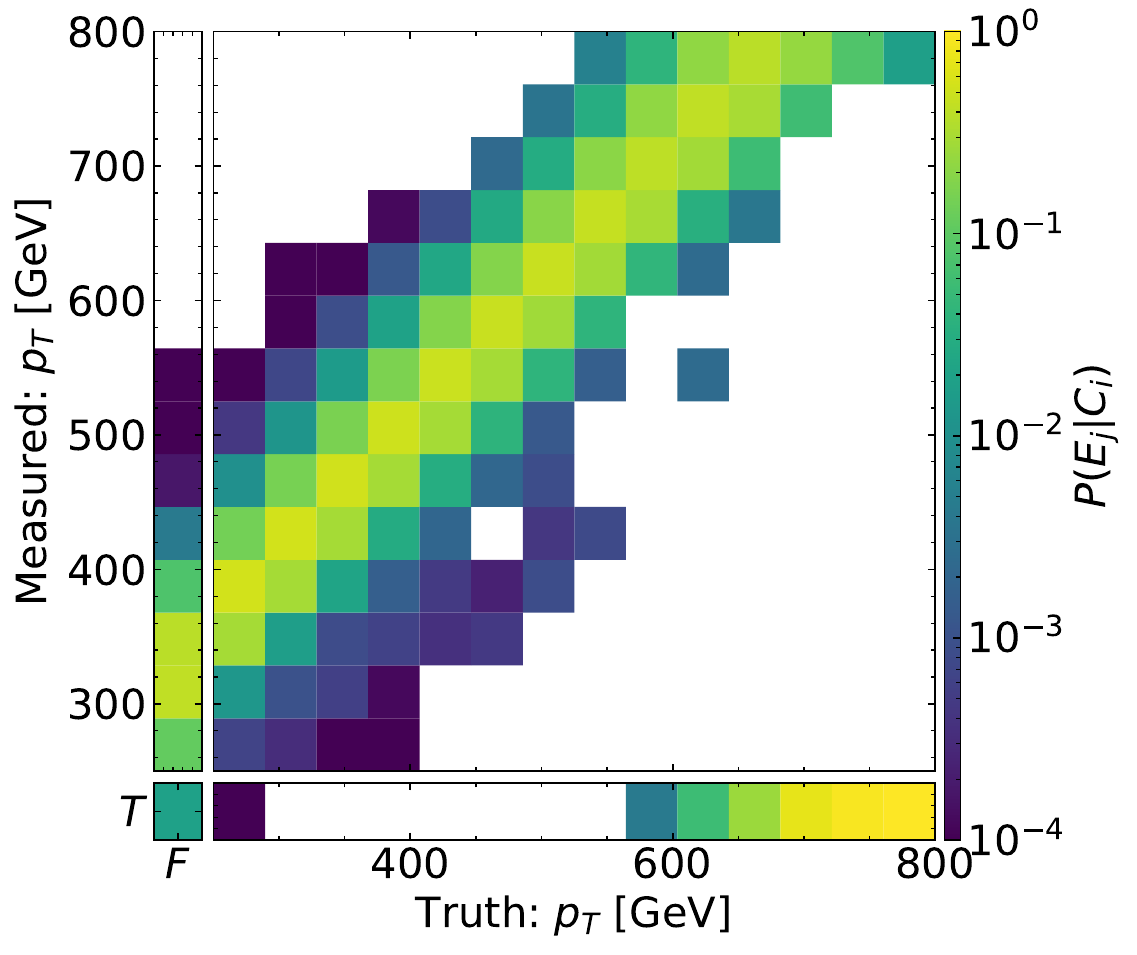}
    \includegraphics[width=0.32\linewidth, page=8]{figs/response.pdf}
    \includegraphics[width=0.32\linewidth, page=15]{figs/response.pdf}
    \includegraphics[width=0.32\linewidth, page=127]{figs/response.pdf}
    \includegraphics[width=0.32\linewidth, page=134]{figs/response.pdf}
    \includegraphics[width=0.32\linewidth, page=141]{figs/response.pdf}
    \includegraphics[width=0.32\linewidth, page=253]{figs/response.pdf}
    \includegraphics[width=0.32\linewidth, page=260]{figs/response.pdf}
    \includegraphics[width=0.32\linewidth, page=267]{figs/response.pdf}
    \caption{The response matrix Eq.~\eqref{eq:RespMX} of truth and measured events with the \texttt{Herwig} (+ Bkg. + \texttt{Delphes}). The small panels show fake and trash.}
    \label{fig:obs_corr}
\end{figure}

As discussed in Eq.~\eqref{eq:Unfolding_basics}, the response matrix captures the probabilistic mapping from cause to effect:
\begin{equation}\label{eq:RespMX}
    p({\rm effect}_j|{\rm cause}_i)=\frac{p({\rm effect}_j,{\rm cause}_i)}{p({\rm cause}_i)}\,,
\end{equation}
where $j,i$ label bins across observables and $p(\_)$ are probability densities. Figure~\ref{fig:obs_corr} presents the response matrix for $p_T,k_{t,g},$ and $\tau_2$ from the \texttt{Herwig} dataset. Our auxiliary files contain plots for all observables. The \texttt{Pythia} dataset follows a very similar response matrix, but not identical. There are also slight differences in the phase space coverage, for instead, \texttt{Pythia} produces larger multiplicities. These differences provide fundamental limitations in the unfolding performance. The upper-left panel shows a significant $p_T$ shift ($\approx100$ GeV) and smearing. The center panel shows correlation in $k_{t,g}$, but with contamination from soft background splittings ($\approx10$ GeV) mistakenly identified as the hardest. The lower-right panel shows both shift and smearing in $\tau_2$. Off-diagonal panels reveal nontrivial correlations among the observables. Compared to multidimensional unfolding, one-dimensional (1D) unfolding ignores these correlations, resulting in biased results and increased uncertainties. Subpanels further indicate fake and trash probabilities. Observable values outside the plotted range are automatically categorized as fake or trash, ensuring that each response matrix column sums to one.

\subsection{Comparison of IBU and OmniFold-HI}
\label{sec:IBUvsOFHI}

We implemented the IBU and OmniFold algorithms from Sec.~\ref{sec:Unfolding}, incorporating our extensions to account for fakes, trash, inefficiencies, and statistical uncertainties. This enhanced version is referred to as OmniFold-HI. We explored several neural network (NN) architectures, by varying hyperparameters such as the number of hidden layers and respective width, activation functions, batch size, number of epochs, and techniques for mitigating overfitting, e.g. early stopping, dropout, and regularization. The NNs in OmniFold-HI are implemented with Keras~\cite{chollet2015keras}, with two fully connected 200 node-wide hidden layers, each with ReLu activation functions, and a sigmoid activation function for the single output. The input size matches the number of considered observables. The NNs are trained using the Adam optimization algorithm, with 20 epochs, and a batch size of 8192. All code used for event generation, detector simulation, dataset preparation, and unfolding algorithms are publicly available in~\cite{OmniFoldHI_GitHub}.

We use \texttt{Pythia} as the generated/simulated, and \texttt{Herwig} as the truth/measured dataset. This choice reflects a realistic analysis workflow and demonstrates the method's robustness to the choice of the event generator. The study of heavy-ion Monte Carlo event generators is left for future work, particularly addressing the challenge of populating adjacent phase space regions, a known issue even in $pp$ analyses~\cite{CMS-PAS-JME-23-001}. For the unfolding, \texttt{Pythia} dataset was the prior and response matrix, and \texttt{Herwig} served as the measurement. The goal was to unfold and predict the \texttt{Herwig} truth distribution.

\begin{figure}
    \centering
    \includegraphics[width=\linewidth]{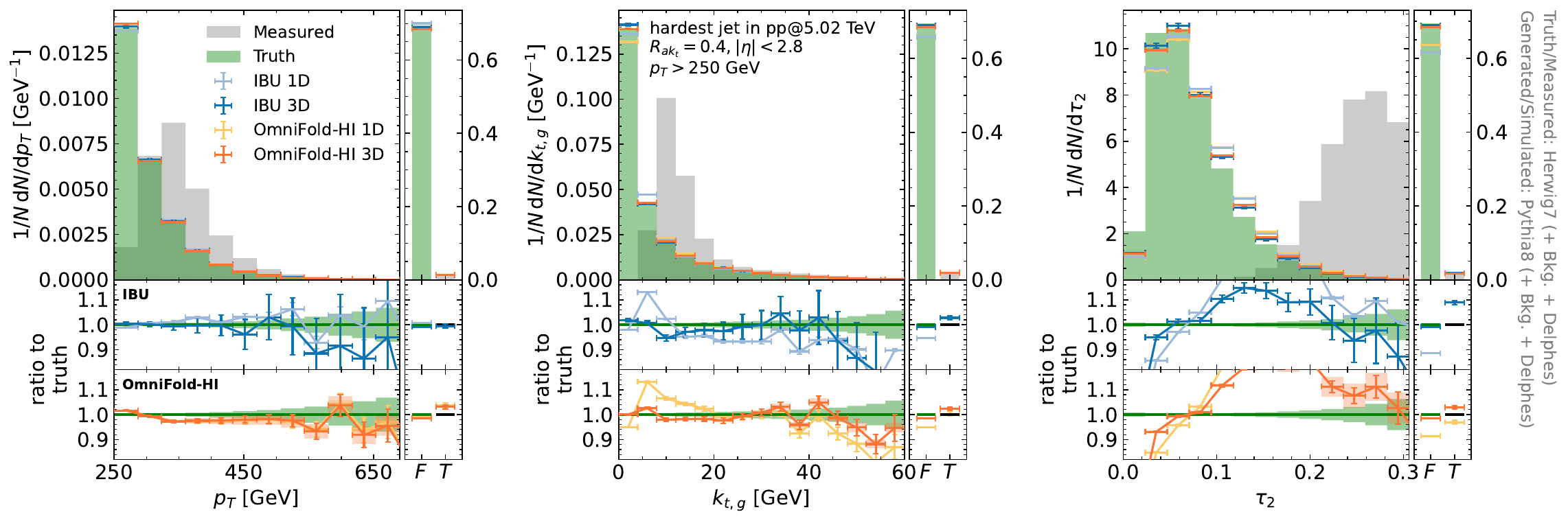}
    \caption{The comparison of the IBU and OmniFold-HI algorithms for 1D and 3D unfolding after three iterations. The 1D unfolding is done for each observables separately. The side panels show the unfolding of trash and fakes.}
    \label{fig:IBU_vs_OF}
\end{figure}

As discussed in Sec.~\ref{sec:Unfolding}, IBU and OmniFold-HI are two distinct implementations of the same iterative unfolding framework. Figure~\ref{fig:IBU_vs_OF} illustrates this by comparing their performance in both one-dimensional (1D) and three-dimensional (3D) unfolding scenarios: specifically, $p_T\to p_T$ (1D), $k_{t,g}\to k_{t,g}$ (1D), $\tau_2\to\tau_2$ (1D), and $\{p_T,k_{t,g},\tau_2\}\to\{p_T,k_{t,g},\tau_2\}$ (3D), after three iterations.

As seen in Fig.~\ref{fig:IBU_vs_OF}, IBU and OmniFold-HI yield very similar results (see ratio panels). They both successfully unfold all observables within 10-20\% accuracy. Notably, the 3D unfolding outperforms 1D due to its ability to capture nontrivial correlations between observables, especially in $k_{t,g}$ and $\tau_2$. The unfolding is not perfect (see e.g. $\tau_2$) which is partially due to remaining prior biases (finite number of iterations), and differences in the event generation and detector simulation (e.g. different response matrix and phase space coverage). We will see that some of these imperfections will vanish when going to higher dimensions. 

The side panels in Fig.~\ref{fig:IBU_vs_OF} show the unfolding of fake and trash probabilities. These are necessary quantities in measuring cross-sections. Since the leading jet $p_T$ serves as the dominant selection criterion, unfolding approaches that incorporate $p_T$ naturally perform better. These panels also demonstrate that OmniFold-HI functions as intended, offering a clear improvement to previous works. For a thorough explanation of uncertainties in Fig.~\ref{fig:IBU_vs_OF}, see the upcoming paragraphs.

\begin{figure}
    \centering
    \includegraphics[width=\linewidth]{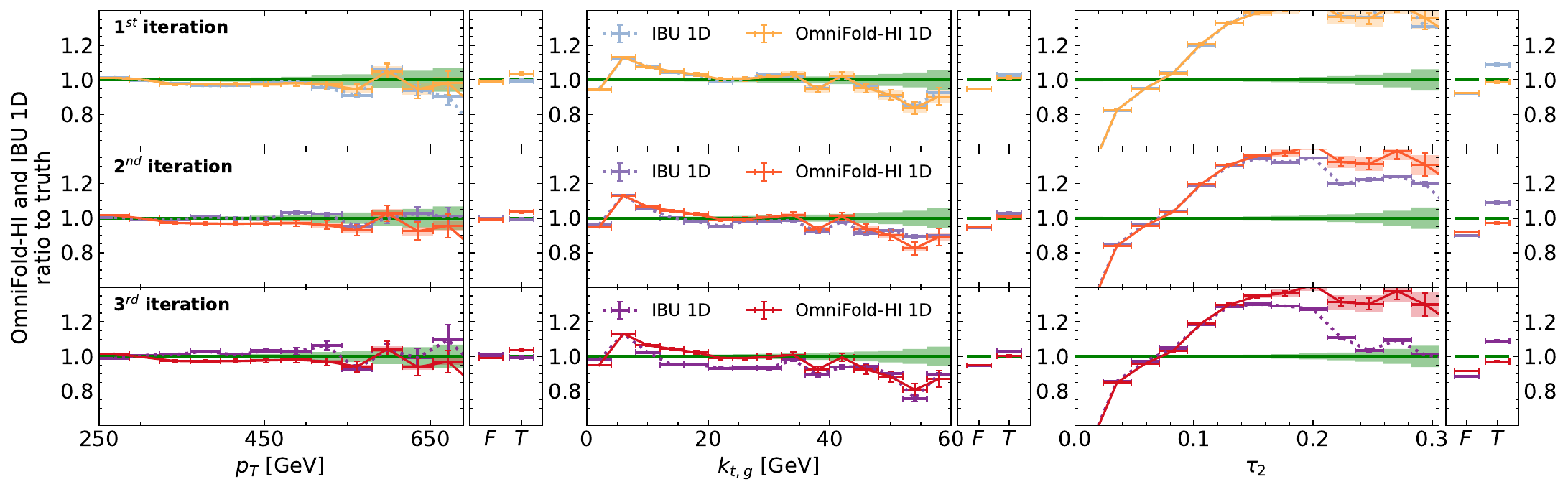}
    \includegraphics[width=\linewidth]{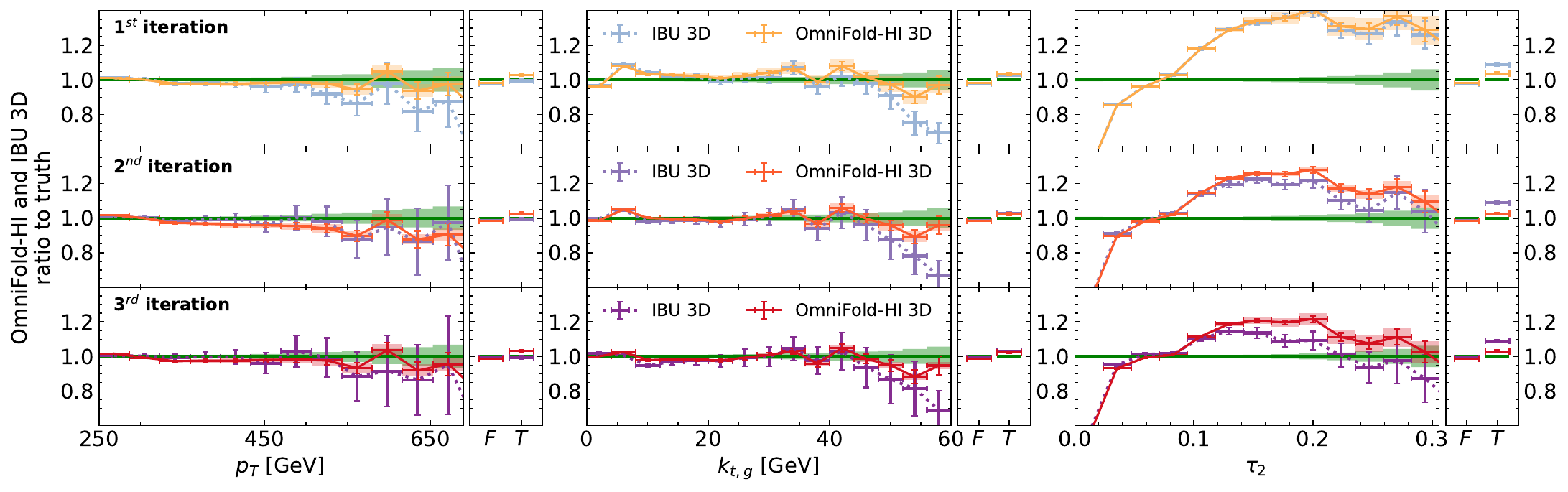}
    \caption{A more detailed comparison of the IBU and OmniFold-HI algorithms for the first three iterations and in different dimensions. See the text for a detailed description of error bars.}
    \label{fig:ofhi_match_ibu_dratio}
\end{figure}

\paragraph{Convergence and uncertainties:}
While IBU and OmniFold-HI are mathematically the same and produce similar results in Fig.~\ref{fig:IBU_vs_OF}, their output is not identical numerically. To further investigate, Fig.~\ref{fig:ofhi_match_ibu_dratio} compares the output of both algorithms over the first three iterations (only ratios are shown). As expected, both methods coincide at the prior and the first iteration. However, small deviations appear in subsequent iterations, likely due to imperfections in the neural network approximations, especially for the statistically less populated regions. The a priori binning of IBU in contrast to the a posteriori binning of the unfolded OmniFold-HI might further emphasize these differences.

Statistical uncertainties are represented with error bars and handled differently for the two methods (for more details see Sec.~\ref{sec:StatUnc}). For IBU, error bars are obtained via bootstrap resampling of the measured dataset: each event is assigned a Poisson weight ($\langle w\rangle=1$), and the unfolding is repeated multiple times. The mean and standard deviation of the resulting distributions yield the error bars. This uncertainty grows with iteration, and often exceeds the statistical uncertainty of the truth distribution (green band). This well-known limitation of IBU is usually mitigated by stopping the iterations early or applying smoothing~\cite{DAGOSTINI1995487,2010arXiv1010.0632D}. This explains the big uncertainties for IBU in Figs.~\ref{fig:IBU_vs_OF} and \ref{fig:ofhi_match_ibu_dratio}, especially for 3D.

In contrast, OmniFold-like NN models do not inherently propagate measurement uncertainties in the same way. To address this, we adopted a new uncertainty estimation method (see Sec.~\ref{sec:StatUnc} and Ref.~\cite{Nachman:2020fff}): a second unfolding step is performed to directly unfold the distribution of uncertainties. For a full discussion on the uncertainty estimation and propagation in OmniFold-HI, see Sec.~\ref{sec:Uncertainties}, and App.~\ref{sec:StatUnc_OFHI}. The resulting error bars in Fig.~\ref{fig:ofhi_match_ibu_dratio} reflect this new method. These uncertainties remain stable across iterations and closely match the statistical uncertainty of the truth distribution. This uncertainty is certainly smaller then the IBU one as the NN training introduces a regulator to prevent overfitting that is not included in IBU. The same uncertainties are shown in Fig.~\ref{fig:IBU_vs_OF}. At this point it is not clear how to recover all IBU uncertainties and we leave it for future work.

To quantify network imperfections, we retrained the networks multiple times with varying initializations. The resulting spread is shown as systematic error boxes in Fig.~\ref{fig:ofhi_match_ibu_dratio}. These uncertainties notably decrease with iterations. A more accurate estimation would also vary the NN hyperparameters. Network and statistical uncertainties combined provide an agreement between IBU and OmniFold-HI. The same uncertainties are also shown in Fig.~\ref{fig:IBU_vs_OF}.

\subsection{Unfolding in higher dimensions}

Having established that OmniFold-HI reproduces the results of IBU we now examine its performance in higher-dimensional settings. A key advantage of OmniFold is its scalability: extending to more dimensions merely involves increasing the input dimensionality of the neural network to match the number of observables, with minimal impact on computational runtime.

\begin{figure}
    \centering
    \includegraphics[width=\linewidth]{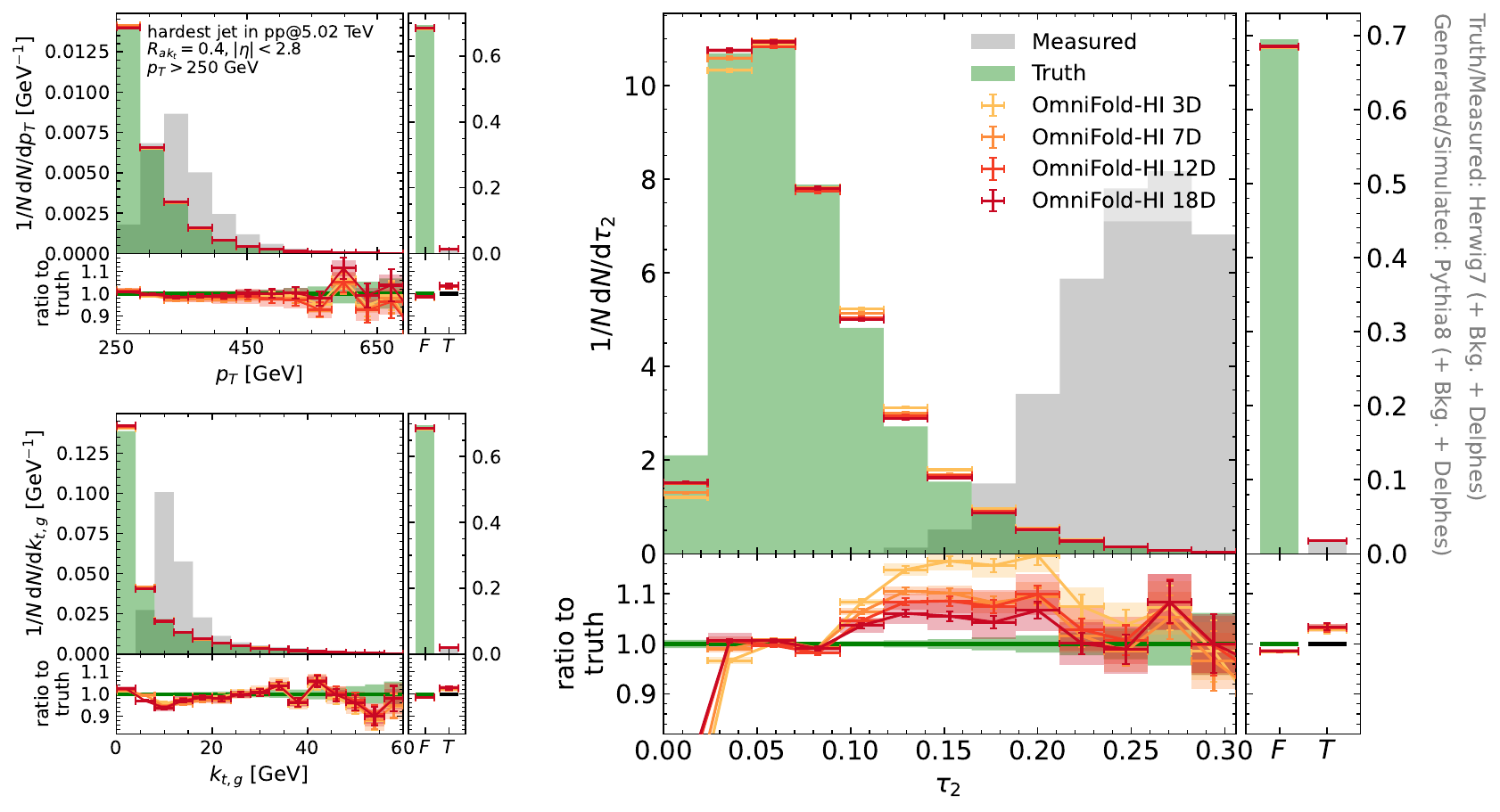}
    \caption{The unfolding of jet observables in higher dimensions by including auxiliary observables. Five unfolding iterations are used. Error bars represent statistical uncertainties, and error boxes network uncertainties.}
    \label{fig:OF_dim}
\end{figure}

Figure~\ref{fig:OF_dim} shows the unfolding performance of the target observables $p_T,k_{t,g}$ and $\tau_2$ in a multi-dimensional context, after five unfolding iterations, that is used from now on. This is achieved by including \textbf{auxiliary observables} to the unfolding input set.\footnote{
    In order: $p_T, k_{t,g}, \tau_2$ (3D), $\eta, \phi, n, m$ (7D), $z_g, \theta_g, p_T^{sd}, m_{sd}, z_{sd}$ (12D), $p_T^{rsd}, n_{rsd}, \tau_1, \tau_3, \tau_4, \tau_5$ (18D).}
While the 3D unfolding already yields strong agreement with truth-level distributions, the inclusion of additional variables leads to further performance gains, most notably for $\tau_2$ and $p_T$. We would like to emphasize that these results include realistic heavy-ion background and full detector simulation and it results in percent level accuracy. Compared to traditional 1D techniques (see Fig.~\ref{fig:IBU_vs_OF}) these are significant improvements. These gains are direct consequences of performing the common iterative Bayesian unfolding in high-dimensions rather than parameterizing a black box deep NN to predict the results. The analysis of other observables (see auxiliary figures) suggests that others, e.g. $n, n_{rsd}$, $m_{sd}$, and $\tau_i$, also benefit substantially from high-dimensional unfolding. In Fig.~\ref{fig:OF_dim}, error bars represent the statistical uncertainty, which were directly unfolded (see Sec.~\ref{sec:StatUnc}); whereas error boxes represent the NN uncertainties that we quantified by retraining the same network with different initializations.

We interpret these gains as: there are subtle differences between MC generators, and so they respond differently to detector effects. By incorporating more observables, these differences are more fully captured, resulting in enhanced unfolding performance. In the very large dimension limit, unfolding should learn that measurements are embedded and detector simulated version of MC events. Hence, Fig.~\ref{fig:OF_dim} highlights a promising strategy: extending the dimensionality of the observable space can significantly improve the unfolding performance and thus reduce unfolding uncertainties of jet analyses. However, we note that systematic uncertainties related to the NN training tend to increase with dimensionality, as the number of parameters to be trained also grows, an expected trade-off. Figure~\ref{fig:OF_dim} shows that in our settings, the gain in the unfolding performance can be larger than the increased systematic uncertainty depending on the observable.

\begin{figure}
    \centering
    \includegraphics[width=0.53\linewidth]{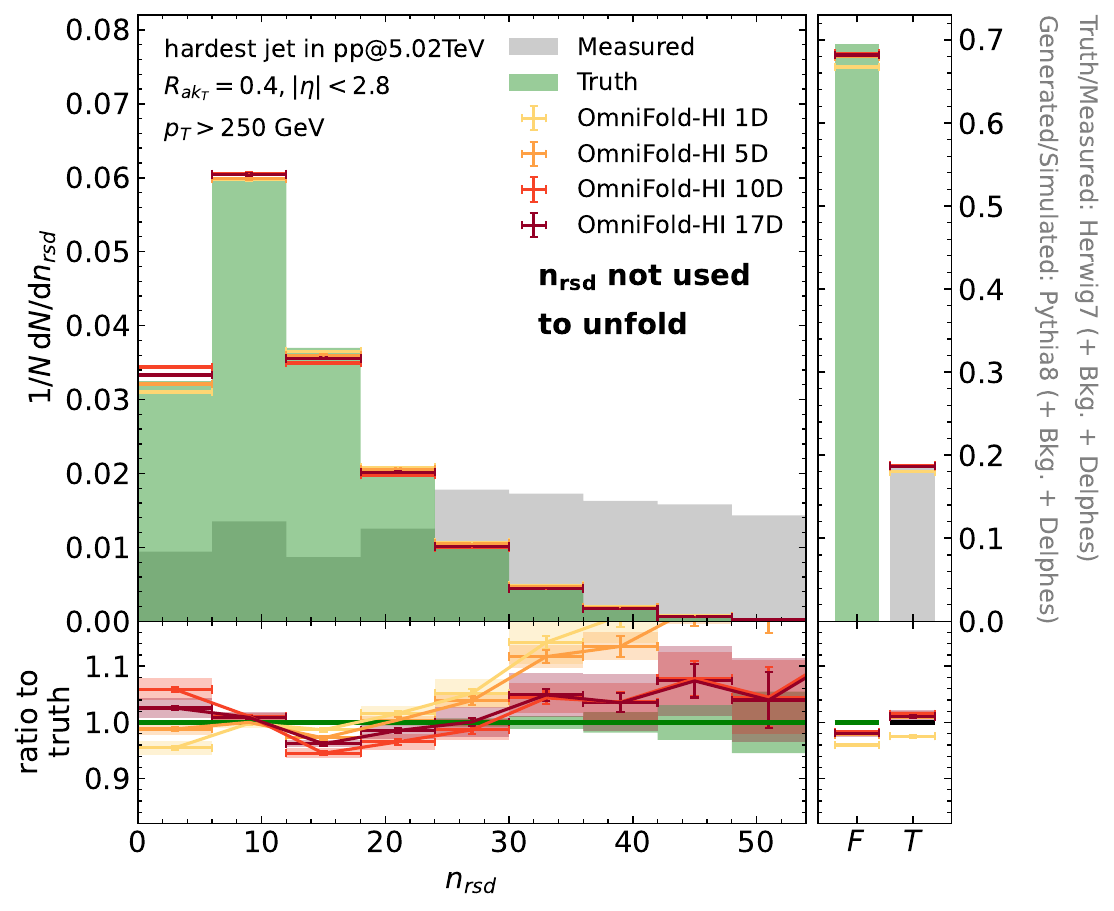}
    \includegraphics[width=0.45\linewidth]{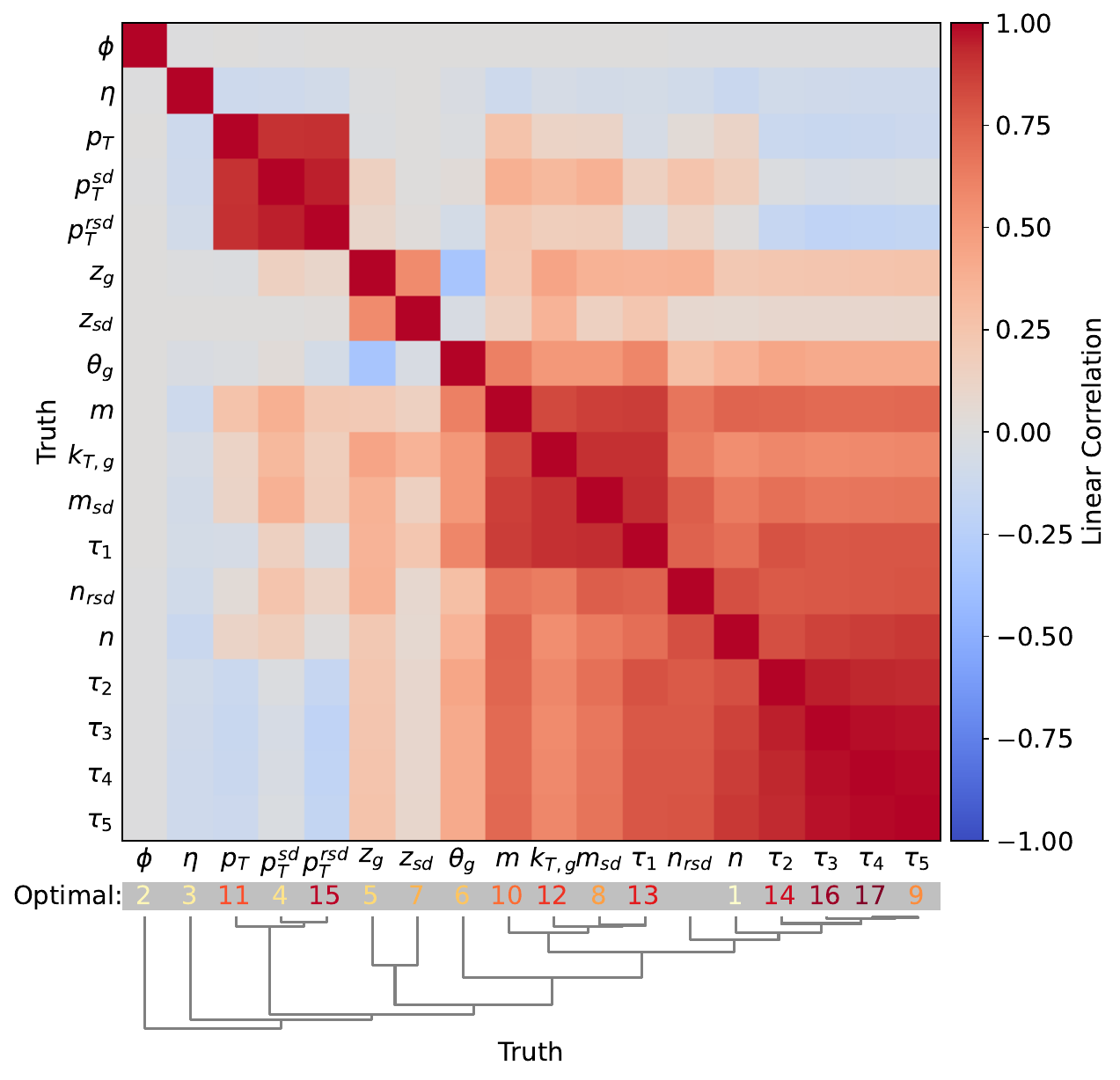}
    \caption{\textit{Left}: Event-level unfolding applied to the recursive Soft Drop multiplicity. Observables were chosen through the optimal unfolding strategy that ensure quick convergence. Uncertainty representation follows Figure~\ref{fig:OF_dim}. \textit{Right:} Linear correlation among observables of \texttt{Herwig} truth events. The clustering tree in the bottom shows groupings among correlated observables which we used for optimal unfolding.}
    \label{fig:OF_dim_evol}
\end{figure}

\paragraph{Event-wide and optimal unfolding:}
Another distinctive strength of OmniFold lies in its use of event weights, enabling the statistical unfolding of the entire event. This allows us to unfold observables which were \textbf{not} included in the training samples. The left panel of Fig.~\ref{fig:OF_dim_evol} illustrates by unfolding the recursive Soft Drop multiplicity $n_{rsd}$, which quantifies the number of hard splittings within jets. Notably, $n_{rsd}$ was not part of the training set. It's unfolding improves markedly with the inclusion of more auxiliary observables, as shown in the ratio panel, reaching a very good agreement with the truth value. These gains arise from capturing nontrivial correlations between observables, which are implicitly captured through the higher-dimensional representation.

This raises the question: how should one select auxiliary observables? To address this, we introduce a strategy, achieves substantial gains without going to extremely high dimensions. To understand better the observable space, the right panel of Fig.~\ref{fig:OF_dim_evol} shows the linear (Pearson) correlation among observables for \texttt{Herwig} truth-level events.\footnote{
    As some correlations are non-linear, one could consider other correlation measures. For example, Kendall's coefficient is such a method for which we thank Jesse Thaler for suggesting that. For our study, it results however in very similar results.
} Here, $n_{rsd}$ shows strong correlation with jet multiplicity $n$ and hard substructure observables like $\tau_i$, while remaining largely uncorrelated with $\eta$, $\phi$, $p_T$, and $z_g$. 

To achieve an optimal selection that covers the entire phase space, we characterize the observable phase space, by clustering observables based on their correlation distance, defined as $d_{ij} = 1 - |{\rm LC}(i,j)|$. The resulting clustering tree is shown in the lower right of Fig.~\ref{fig:OF_dim_evol}. Strongly correlated observables like $\tau_i$ group together, while observables like $\phi$ and $\eta$ stand alone. We use these clusters to optimally select the auxiliary observables. For a given unfolding dimension, we resolve the same number of clusters.\footnote{Non-grouped observables count as individual clusters.} From each of these clusters we choose the most correlated observable to unfold relative to the target observable ($n_{rsd}$ for us). The resulted procedure is denoted with the numbers in Fig.~\ref{fig:OF_dim_evol}.\footnote{
    In order: $n$ (1D), $\phi,\eta,p_T^{sd},z_{g}$ (5D), $\theta_g,z_{sd},m_{sd},\tau_5,m$ (10D), $p_T, k_{t,g},\tau_1,\tau_2,p_T^{rsd}$ (15D) $\tau_3,\tau_4$ (17D).
}
This method ensures a quick convergence with increased dimensions in the left of Fig.~\ref{fig:OF_dim_evol}. Going to 17D from 10D doesn't further improve the performance fro this particular observable. In our studies we also tried to select observables based on the correlation strength with $n_{rsd}$ resulting in a slower convergence for increasing dimensions (not shown).

\subsection{Unfolding with calibration}
\label{sec:Unf_w_Calib}

In the previous sections, we unfolded directly from the reconstruction to the truth level, bypassing the usual intermediate step of jet calibration. Calibration intends to bring the reconstructed jet momentum closer to its true value. In this section, we argue that unfolding directly from the reconstruction level yields improved performance by preserving important correlations between observables.

During calibration, the reconstructed jet 4-momentum is corrected via jet energy scaling (JES). This process involves auxiliary quantities like the event activity and several reconstructed jet properties, and it uses parametrized formulas to correct the reconstructed jet momentum to match the truth on average. Fluctuations around the mean are also corrected, as the measurement typically has larger variations. Fluctuations in the data are parametrized using the jet energy resolution (JER) that is applied to smear the simulated momenta to better match the variance of the data.
The JER is often parametrized using simplified formulas that overlook the complex correlations between different jet observables. For example, larger jet momentum or mass also involves larger multiplicity and a harder substructure, a correlation that JER smears out even when JER and JES uncertainties are included. In what follows, we show that such a smearing procedure reduces physical correlations, worsening the unfolding performance in higher dimensions.

We apply a simplified calibration procedure to our setup, in which we align the jet momenta in simulation with those in the measurement at the level of their mean and variance. Ideally, one would employ the internal calibration routines used by experimental collaborations; however, these are not available to us. Instead, we implement the following procedure:
\begin{enumerate}
    \item Analogous to the JES correction, we evaluate $S_i=\langle p^\mu\rangle_{{\rm gen},i}/\langle p^\mu\rangle_{{\rm sim},i}$ for different $p^\mu_{\rm sim}\in [p^\mu_i,p^\mu_i+\Delta^\mu]$ bins. Then, the corrected momenta are $p^\mu_{{\rm sim},i}\mapsto S_i\cdot p^\mu_{{\rm sim},i}$, and $p^\mu_{{\rm meas},i}\mapsto S_i\cdot p^\mu_{{\rm meas},i}$. This results a matching between the average ``gen'' and ``sim'' for every $p_i^\mu$ bin and the corresponding correction to the measurement.
    \item Analogous to the JER correction, we evaluate $\sigma^2_{{\rm mc},i}={\rm Var}(p^\mu_{{\rm gen},i}/p^\mu_{{\rm sim},i})$ for the same bins. The $\sigma_{{\rm meas},i}$ is measured experimentally e.g. in boson-jet correlations. Instead, we assume simply $\sigma_{{\rm meas},i}^2=2\sigma_{{\rm mc},i}^2$. As typically $\sigma_{\rm meas}>\sigma_{\rm mc}$, the simulation needs to be smeared to match the variation. We use a normal distribution, $p^\mu_{{\rm sim},i}\to S_i\cdot p^\mu_{{\rm sim},i} + \mathcal{N}(\mu=0,\sigma^2=\sigma^2_{{\rm meas},i}-\sigma^2_{{\rm mc},i})$. Therefore, ``sim'' and ``meas'' variances match each other for every bin.
\end{enumerate}
After calibration, the first and second moments satisfy $\langle{\rm Calib}(p^\mu)\rangle_{{\rm sim},i}=\langle p^\mu\rangle_{{\rm gen},i}$, and ${\rm Var}({\rm Calib}(p^\mu))_{{\rm sim},i}={\rm Var}({\rm Calib}(p^\mu))_{{\rm meas},i}$. Crucially, substructure observables stayed the reconstructed ones, and the smearing is clearly one-dimensional, decorrelating observables. Uncertainties to the parametrized distributions are neglected. The unfolding is then performed using the calibrated data and smeared simulation.\footnote{
    We calibrated only $p_T$ and neglected $m$. $\eta$ and $\phi$ are also not calibrated as they are typically matched geometrically in analyses, a method we did not use.}
This simplified method resembles typical calibration procedures at CMS~\cite{CMS:2011shu,CMS:2016lmd,CMS:2017ehl,CMS:2018dqf}, and ATLAS~\cite{ATLAS:2019oxp,ATLAS:2020cli,ATLAS:2023tyv}. Therefore, cross-correlations among jet 4-momentum, and jet momentum and substructure, visible as off-diagonal panels in Fig.~\ref{fig:obs_corr}, are smeared in this procedure.

\begin{figure}
    \centering
    \includegraphics[width=\linewidth]{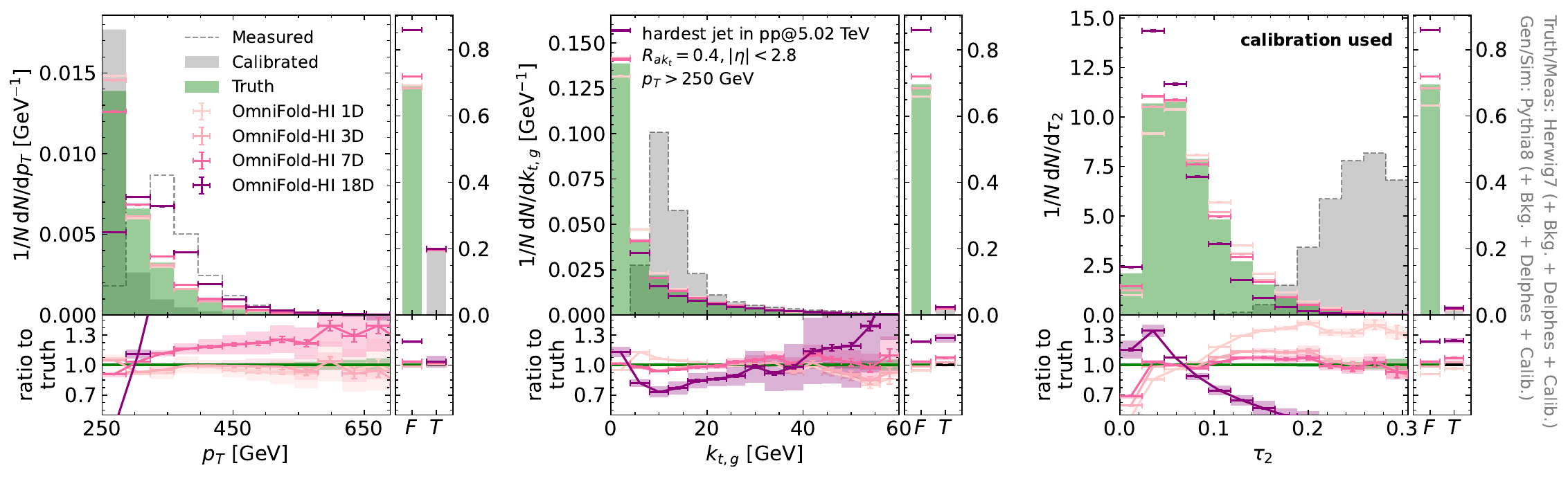}
    \caption{The unfolding of jet observables in various dimensions after applying jet calibration (JES + JER). Calibration has clearly worsen the performance compared to Fig.~\ref{fig:OF_dim}.  Uncertainty representation follows Fig.~\ref{fig:OF_dim}. Note the larger scale used in the ratio to truth panels, compared with the one used in all the other figures.}
    \label{fig:OF_dim_cal}
\end{figure}

Unfolding under such assumptions is shown in Fig.~\ref{fig:OF_dim_cal}. The calibrated $p_T$ distribution is much closer to the truth, while $k_{t,g}$ and $\tau_2$ are unchanged. Note that, in Fig.~\ref{fig:OF_dim_cal}, the calibrated $p_T$ distribution is slightly lower because our simple calibration increased the number of trash counts. The one-dimensional unfolding of $p_T$ still yields accurate results, while $k_{t,g}$ and $\tau_2$ hasn't changed. Higher dimensional unfolding performs gradually worse compared to Fig.~\ref{fig:OF_dim} as due to smearing, correlations are no longer identical in simulated and measured data. The unfolding performance of fake and trash events are also largely affected, and these quantities are essential in measuring total cross-sections.

This comparison supports our argument that unfolding from the reconstruction level, without prior calibration, provides superior performance. As a practical compromise, selecting jets using calibrated quantities while unfolding using reconstructed quantities can balance the benefits of both approaches. Furthermore, calibration provides a natural source of auxiliary observables for unfolding, since it already involves several event-level quantities that are used during JES but not reported in the final results. Integrating jet calibration and unfolding to a single step would result in further gains in performance.

\subsection{Unfolding with particle-level background subtraction}

In dense environments, a significant portion of jet calibration corrections arises from the removal of pileup or underlying event contributions. In this section, we demonstrate a method (commonly used in the heavy-ion context) that performs such subtraction while preserving correlations among observables. This technique operates at the particle level, effectively integrating background subtraction into the jet reconstruction process itself. We show that this approach reproduces the benchmark unfolding performance and outperforms traditional techniques from Sec.~\ref{sec:Unf_w_Calib}.

The Iterative Constituent Subtraction (ICS) method~\cite{Berta:2019hnj} is designed to remove background particles from events by estimating and subtracting the local transverse momentum ($p_T$) density in the $\eta$–$\phi$ plane. It does this by inserting ghost particles with negative momenta, based on the estimated background density. Importantly, ICS applies corrections to the full event rather than to isolated jet quantities, helping to retain non-trivial correlations between observables.

We applied ICS to our measured samples. To estimate the background, we used the \texttt{GridMedianBackgroundEstimator} from \texttt{fastjet}, with a grid size of 0.5 and an active ghost area of 0.01. The subtraction was performed in two iterations, with maximum distances of $(0.1, 0.2)$ and $\alpha = 1$, using the \texttt{fjcontrib} package interfaced with \texttt{Delphes}. We then unfolded these background-subtracted samples to the truth level.

\begin{figure}
    \centering
    \includegraphics[width=\linewidth]{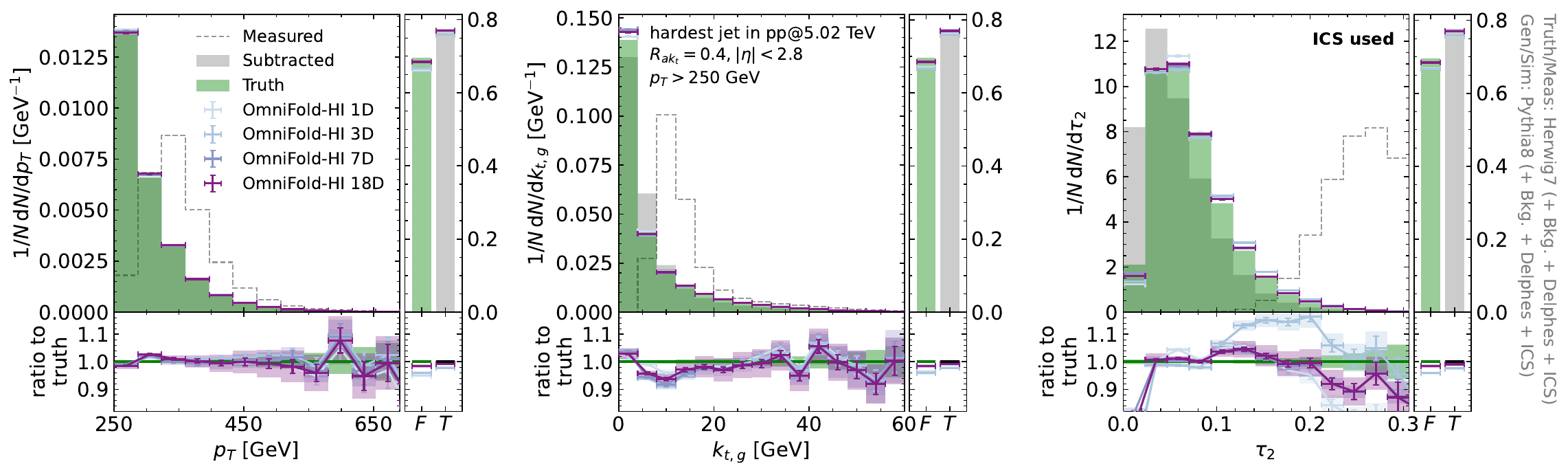}
    \caption{Unfolding of jet observables in multiple dimensions after applying particle-level background subtraction to the measured distributions. The application of particle-level jet corrections preserves the baseline performance observed in Fig.~\ref{fig:OF_dim}. Uncertainty representation follows Fig.~\ref{fig:OF_dim}.}
    \label{fig:OF_dim_ics}
\end{figure}

Figure~\ref{fig:OF_dim_ics} presents the multidimensional unfolding results for our key observables. After applying ICS, the measured distributions (gray histograms) are significantly closer to the truth-level distributions (green histograms) across all variables. The contribution from fake signals did not change, while the number of trash jets falling below the 250 GeV threshold increased due to background removal. When focusing on the 1D unfolding results, we observe clear improvements compared to our earlier results without background subtraction (see Fig.~\ref{fig:IBU_vs_OF}). This improvement is the result of unfolding more similar distributions. Furthermore, we see a good agreement in higher dimensions relative to our previous benchmarks in Fig.~\ref{fig:OF_dim}. This suggests that crucial correlations among observables are maintained even after the subtraction procedure. Therefore, the subtraction did not increase performance but rather maintained it across higher dimensions. 

In large backgrounds, calibration mainly corrects the contamination coming from background particles. The ICS illustrates a particle-level, event-by-event subtraction method that, applied to reconstructed jets, successfully approximates most truth-level jet quantities, while maintaining correlations among them. When unfolding these subtracted jets to the truth level, we recover the standard unfolding performance and thus we outperform traditional techniques (Sec.~\ref{sec:Unf_w_Calib}) where correlations are not correctly accounted for.

\section{Conclusions and outlook}
\label{sec:conclusions}

In this paper, we reviewed the Iterative Bayesian unfolding (IBU) and OmniFold algorithms, showing that they are equivalent and maximum likelihood estimators. We provided a rigorous derivation of the OmniFold algorithm, highlighting that its neural network (NN) components serve as abstract representations of histogram (or likelihood) ratios, and we provided physical meaning to its loss functions. These insights challenge the common perception of neural networks as black boxes. Additionally, we introduced OmniFold-HI, which takes into account background, fake signal, detector acceptance, efficiency losses, and uncertainties, which are essential considerations in heavy-ion and high-luminosity phenomenology. Our extensions rely on mathematical formulas, free of additional NNs, and access uncertainties, going beyond the previous extension Ref.~\cite{Andreassen:2021zzk}. In this setup, we showed that IBU and Omnifold-HI are equivalent to the expectation-maximization algorithm, and we established, their numerical equivalence.

The most important contribution of this work is the introduction of auxiliary observables into the unfolding process, resulting in significantly improved unfolding performance on the target observables. This gain is the result of the consistent treatment of correlations among observables. To demonstrate this, we performed a realistic analysis by unfolding detector effects in jet observables in a heavy-ion background, reaching up to 18 dimensions. We observed clear gains in the performance. The introduction of OmniFold-HI was necessary to perform unfolding in such high dimensions.

To choose auxiliary observables, we introduced a new technique that maps the event-space based on the correlation among observables. Only a few optimal quantities are enough to maintain the gains found in high dimensions. We demonstrated this by unfolding a jet substructure observable that was not included in the training process. Our findings shows good unfolding convergence in high dimensions, and are useful for future event-level unfolding, and in cases where resources (e.g. MC samples, computing power) are limited. 

Finally, we proposed a novel strategy in which jet calibration (together with background subtraction) is incorporated directly into the unfolding. We studied this in the presence of large, fluctuating backgrounds. This includes the increased pile-up at high-luminosities, and the underlying event in heavy-ion collisions. This approach fits well with the use of auxiliary observables as calibration uses several reconstructed quantities. Furthermore, it supports the idea of merging calibration and unfolding into a unified step. This method reduces systematic uncertainties through the consistent treatment of correlations among measured quantities. In practice, this would involve unfolding reconstructed (rather than calibrated) observables back to truth-level quantities. Similar approaches are already used on smaller scales, e.g. in ATLAS's residual pile-up correction in proton collisions~\cite{ATLAS:2023tyv}, and in heavy-ion jet measurements by ALICE~\cite{ALICE:2013dpt,ALICE:2015mjv,Haake:2018hqn,ALICE:2023waz}. We demonstrated these by embedding Monte Carlo events into a heavy-ion background with detector distortions and unfolding jet observables directly back to the generated level. We compared this method to a more traditional technique that involves a simplified calibration first. The new method outperforms the traditional one as the latter neglects correlations through the usage of jet energy resolution. Alternatively, we introduced a particle-level subtraction technique that is able to maintain all necessary correlations and could be used to parametrize calibration in the future.

Our new techniques are beneficial across all collider analyses. The code and data used in this study are available in \cite{OmniFoldHI_GitHub}. To our knowledge, this is the first work to apply auxiliary observables, machine learning-based techniques, and to integrate calibration and background subtraction in the context of unfolding in dense collisions. We provided the most detailed study of ML uncertainty propagation in the unfolding context, and showed that traditional calibration techniques have to be reexamined for high-dimensional unfolding.

As an outlook, future research could explore extending our unfolding framework to even higher dimensions using Deep Sets~\cite{Komiske:2018cqr,Andreassen:2019cjw,Komiske:2022vxg,Ba:2023hix,Athanasakos:2023fhq} instead of a finite set of observables. This would allow a more robust unfolding of the entire event. Unfolding directly at the particle level is also possible by following~\cite{Maier:2021ymx,Quetant:2024ftg,Algren:2024bqw}, enabling event-by-event generative unfolding using techniques from~\cite{Bellagente:2020piv,Backes:2022sph,Backes:2023ixi,Diefenbacher:2023wec,Torbunov:2024iki}. We also emphasize that unfolding, background subtraction, fast detector simulation, etc. often share similar methodological frameworks~\cite{Steffanic:2023cyx,Mengel:2024fcl,Stewart:2024mkx,Bierlich:2024xzg}, as these are all simulation-based (likelihood-free) inferences~\cite{2020PNAS..11730055C}. Finally, a systematic comparison of different heavy-ion MCs, background models, calibration methods, unfolding frameworks~\cite{Adye:2011gm,Brenner:2019lmf,Huetsch:2024quz,Milton:2025mug}, and likelihood-ratio learners~\cite{Nachman:2021yvi,Rizvi:2023mws} is available for proton collisions and extending them to the heavy-ion context would provide valuable insights for the community.

\acknowledgments
We greatly appreciate discussions with Hannah Bossi, Raymond Ehlers, Nathan Huetsch, Benjamin Nachman, Martin Spousta, Jesse Thaler, Marta Verweij, and Vangelis Vladimirov. The work is supported by DFG through Emmy Noether Programme (project number 496831614)  (A.T), and through CRC 1225 ISOQUANT (project number 27381115) (A.T).

\appendix
\section{Statistical uncertainty in OmniFold-HI}
\label{sec:StatUnc_OFHI}

This appendix provides illustrations for the conclusions made in Sec.~\ref{sec:StatUnc}. Here, we compare the statistical uncertainties obtained using different uncertainty estimation methods. Below, we describe how the statistical uncertainty is estimated via Poisson bootstrap, and we refer to Sec.~\ref{sec:StatUnc} for the alternative, \textbf{direct uncertainty unfolding} method.

In the bootstrap method, unfolding is repeated several times using different subsets of the measurement. The final unfolded result is the average and variance of these repetitions. The same bootstrapping method is applied to estimate the statistical uncertainty of the measurement and truth distribution. We consider Poisson bootstrapping and obtain the measured subsets in two different ways:
\begin{itemize}
    \item[] \textbf{Poisson reweighting:} Measured events are assigned with random Poisson weights ($\langle w\rangle=1$) and the unfolding is done on these new weighted events. The number of events does not change;
    \item[] \textbf{Poisson resampling:} Measured events are resampled following their Poisson weights. The unfolding is done on these new sets, without weights, and the number of events is the same as the original set.
\end{itemize}

\begin{figure}
    \centering
    \includegraphics[width=\linewidth]{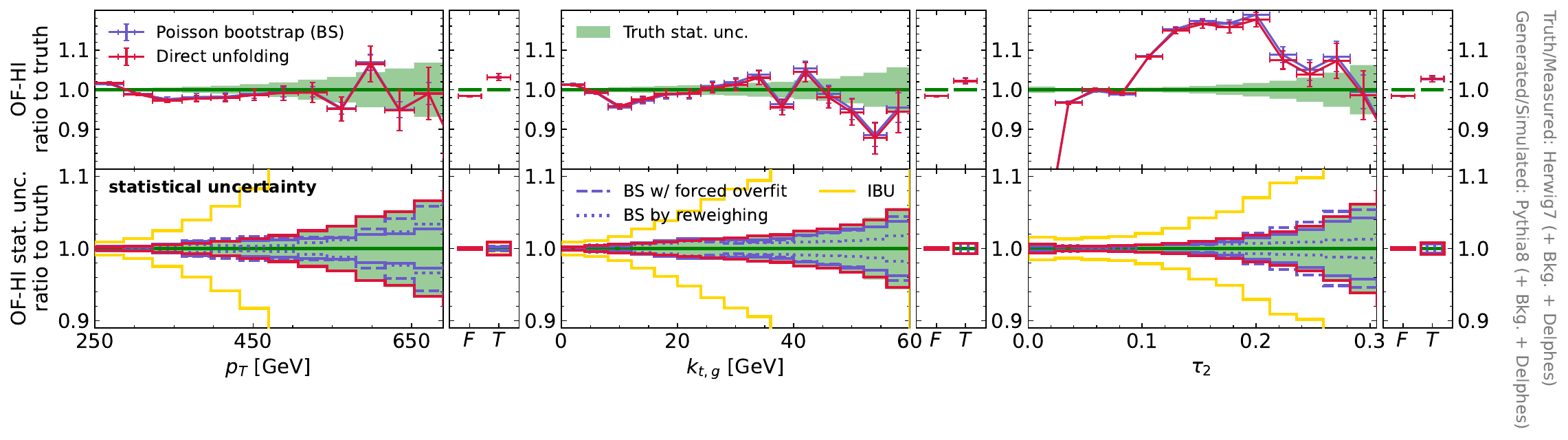}
    \caption{Statistical uncertainties in 3D OmniFold-HI unfolding. We compare different statistical uncertainty estimations: Direct unfolding of statistical uncertainty; and Poisson bootstrapping. Different implementations of Poisson bootstrapping are explored. See text for further details. The IBU uncertainty estimation is also shown as a reference.}
    \label{fig:omnifoldHI_closure_stat}
\end{figure}

Focusing on the nominal unfolded result, unlike bootstrapping, the direct unfolding method does not rely on the use of replicas, so the result of the unfolding bypasses the averaging step. Therefore, different methods can result in different nominal results. In Figure~\ref{fig:omnifoldHI_closure_stat} we show OmniFold-HI for five iterations of the 3D unfolding (also used in Sec.~\ref{sec:IBUvsOFHI}). The upper panels show that the Poisson bootstrap (using resampling) and direct unfolding overlap within their respective uncertainties.

Turning now to statistical uncertainties, the lower panels in Fig.~\ref{fig:omnifoldHI_closure_stat} show that Poisson bootstrap using resampling (purple solid line) underestimates the statistical uncertainty of the truth distribution (green band) that is also estimated through Poisson bootstrap. We attribute this to the NN's resilience to statistical fluctuations. We verified this by forcing overfitting during training. The forced overfitting partially recovers these fluctuations (dashed lines). For a correct uncertainty propagation, one would have to ``perfectly overfit'' the NN at every step.

For reference, we also present the uncertainty estimation obtained through Poisson bootstrapping via reweighting.  This method follows what is usually done in physics analyses with IBU and, to our best knowledge, in previous OmniFold works. The resulting statistical uncertainty (dotted lines in Fig.~\ref{fig:omnifoldHI_closure_stat}) significantly underestimates the statistical uncertainty of the truth distribution, specially when compared to the other methods. This suggests that OmniFold is less sensitive to the use of reweighing while keeping the same events, than to using event resampling, despite the similarities of the subsets. Overall, the best agreement is observed when the statistical uncertainty is directly unfolded, using the algorithm described by Eq.~\eqref{eq:OFHI_algorithm_square}.

Finally, the lower panels of Fig.~\ref{fig:omnifoldHI_closure_stat} show the statistical uncertainty obtained with 3D IBU using Poisson bootstrap with weights. This results in the largest statistical uncertainties. We explain this with the fact that we did not employ any specific regularization (e.g. smoothing) between steps other than the finite iteration number five. In contrast, the prevention of overfitting in NNs should translate to some effective smoothing of histograms in IBU. Therefore, in it's current form, OmniFold always results in smaller statistical uncertainties than IBU. Even though IBU and OmniFold-HI are equivalent algorithms, the way they propagate statistical fluctuations is not trivial, and we leave such a study for future work. 

\bibliography{refs_vac,refs_med}

\end{document}